# Transparentize A Shallow Cryosphere: High-Resolution Subsurface Imaging using UAV-Borne GPR

A review and prospective

WENHAO LUO, TONG HAO, QIAN MA, CHEN LV, AND ZHIYI CAO

Accurate characterization of shallow cryosphere subsurface structures is critical for understanding snow and ice dynamics, evaluating environmental hazards, and informing climate-related decision making. Recent advances in unmanned aerial vehicles (UAVs) and compact radar instrumentation have enabled UAV-borne ground penetrating radar (GPR) as a non-contact alternative for subsurface sensing in these environments. By combining the safety-constrained mobility of unmanned aerial platforms with the penetration capability of radar, UAV-borne GPR facilitates rapid, wide-coverage and flexible surveys over snow and ice surfaces in cryosphere that are often inaccessible to traditional ground-based methods. This contribution reviews recent progress in UAV-borne GPR for cryosphere investigations, encompassing system architectures, operational strategies, and representative deployment practices. Furthermore, key data-processing methodologies are summarized for their roles in enhancing subsurface imaging and reconstruction fidelity. A case study from Mochou Lake in East Antarctica, illustrates these capabilities, yielding an ice-thickness estimation error of only 0.03 m when benchmarked against drilling measurements. Finally, the remaining challenges, particularly those associated with scanning coverage, real-time data processing, and detection accuracy, are discussed for future cryosphere applications.

## INTRODUCTION

The cryosphere, comprising snowpack, ice forms (including glacier, sea ice, iceberg, etc.), and permafrost, constitutes a critical component of the Earth's climate system [1]. It governs surface energy balance through high albedo, regulates global hydrological and biogeochemical cycles, and modulates sea-level variability via ice mass exchange. Snowpack and glacier dynamics control seasonal water availability [2], whereas the stability of sea ice, ice shelves, and icebergs influences ocean circulation and polar ecosystems [3]. Permafrost and buried ice serve as major carbon and methane reservoirs, whose degradation under warming conditions can amplify climate feedbacks [4]. Accurate characterization of the structure, extent, and temporal evolution of shallow cryosphere is thus essential for assessing climate sensitivity and predicting environmental change in polar and alpine regions.

Conventional cryosphere investigations rely primarily on satellite, ground-based, and airborne platforms. Optical and radar satellites provide wide spatial and temporal coverage, enabling large-scale monitoring of glacier and ice-sheet evolution [5]. However, satellite data are constrained by coarse spatial resolution, high acquisition costs, and weather-related interference, while their revisit cycles limit temporal flexibility. Ground-based systems offer high accuracy but have limited accessibility in hazardous terrains such as crevassed or unstable ice surfaces [6]. Airborne surveys provide valuable regional observations but are restricted by operational cost, logistics, and safety considerations [7].

Unmanned aerial vehicles (UAVs), or unmanned aerial systems (UASs), overcome many limitations of conventional platforms by enabling flexible, low-cost, and high-resolution data acquisition in remote or hazardous regions [8]. UAVs support both visual line-of-sight (VLOS) and beyond-visual-line-of-sight (BVLOS) operations. Recent advances in BVLOS enables coverage of larger and more inaccessible areas, acquisition of higher-resolution and more frequent datasets, reduced human footprint, and enhanced flying safety [9]. These capabilities allow for multifaceted environmental analysis in challenging cryosphere environments and enable UAVs a powerful tool for high-resolution monitoring, mapping, and subsurface investigations in cryosphere.

Building upon these advances, UAVs were employed for the first time in Antarctic scientific research during the 17th Italian Antarctic Expedition in 2004 [10]. Since then, UAV technology has rapidly diversified, supporting a range of geophysical sensors, including electromagnetic (EM) [11], magnetic [12], gamma-ray [13], gravity [14], LiDAR [15], sonar [16], and radar systems. Early UAV-mounted radar platforms utilized high-frequency Ka-band (24–40 GHz), X-band (8–12 GHz), and C-band (4–8 GHz) antennas [17], [18], demonstrating the feasibility of aerial radar imaging, though their high operating frequencies limited ground penetration and precluded subsurface applications.

However, subsurface properties are critical for understanding cryosphere dynamics and need to be further studied. A major source of uncertainty in understanding cryosphere evolution lies in the insufficient knowledge of shallow subsurface and deep basal conditions beneath ice sheets, and permafrost [1]. Their thickness, internal and basal structures critically control ice dynamics, mass balance, and hydrological connectivity. The Earth's polar regions constitute the largest and most dynamic domains of the global cryosphere. These regions are undergoing rapid transformation, with pronounced declines in ice extent and thickness, as well as rapid thinning along ice-sheet margins [3]. Mass loss from the Greenland ice sheet accelerated markedly between the early 1990s and late 2000s, substantially increasing its contribution to global sea-level rise [19]. Similar losses have been observed in Antarctica, where ongoing imbalance continues to reshape the cryosphere system. Although complete ice sheet disintegration is not imminent, persistent grounding zone retreat already drives significant sea-level rise, threatening low-lying coastal regions. The Intergovernmental Panel on Climate Change (IPCC) projects up to approximately one meter of sea-level rise by 2100, though higher values remain possible [20]. Despite advances in ice-sheet modeling enabled by improved mechanics and refined bed topography, large uncertainties persist, particularly in constraining upper-end projections [21]. Improved observations of ice thickness and basal conditions are critical for reducing model uncertainties and strengthening future sea-level projections.[1].

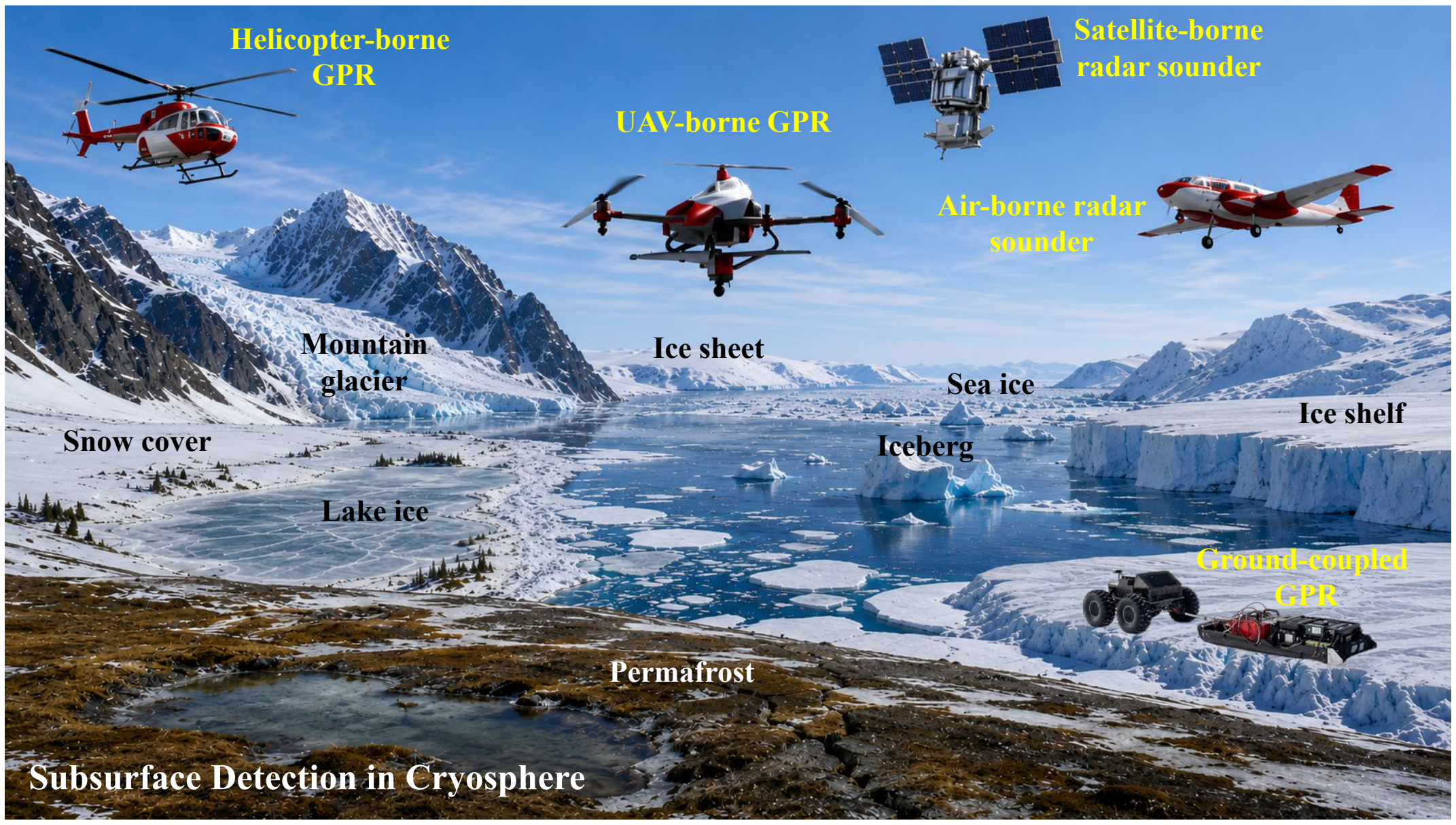

**Fig. 1.** Various GPR systems used in cryosphere subsurface research.

Ground-penetrating radar (GPR) has emerged as one of the most widely adopted technologies for subsurface sensing and imaging, which has been demonstrated to outperform alternative sensing techniques through comparative studies [22], owing to its deep penetration, high resolution, and ability to detect both metallic and non-metallic buried targets. This capability has enabled its use in a broad range of applications, including civil engineering assessments, cultural heritage exploration, and military and public safety operations [23], [24]. In recent years, researchers have also extended GPR applications to cryosphere investigations, such as ice and snow thickness estimation, where ground-coupled systems and fixed-wing airborne platforms remain the predominant approaches [23], [25].

The characterization of subsurface features in cryosphere environments is increasingly critical, and UAV-borne sensing offers new opportunities for high-resolution mapping. This review surveys recent advances in UAV-borne GPR for cryosphere detection and characterization. While prior reviews have considered UAV-GPR integration [25], [26], but they mainly provide broad overviews of UAV–GPR integration across diverse application domains, while cryosphere-related studies are typically discussed only briefly within wider thematic scopes, a focused assessment of its applications in cryospheric settings is still missing. In contrast, this review presents a cryosphere-oriented, system-level assessment framework focusing on cryosphere-specific electromagnetic propagation effects, operational constraints in harsh cryospheric environments, and quantitative mission-design methodologies for UAV-borne GPR in icy media. Specifically, Section IV introduces a structured classification framework for snowpack, ice-form, and permafrost investigations, enabling consistent cross-comparison of system configurations, operating frequencies, and sensing performance across different frozen media. Particular emphasis is placed on permafrost-related signal attenuation barriers, which remain rarely discussed in existing UAV-borne GPR reviews [25], [26]. Beyond critically analyzing the technical, logistical, and environmental factors limiting UAV-borne deployment in permafrost environments, the review further proposes application-driven mitigation strategies. In addition, mission-design heuristics are established by linking UAV flight parameters and radar configurations to target dielectric contrast, penetration depth, and lateral scale. These system–target co-design relationships, informed by both conventional GPR theory and airborne cryospheric radar sounding practices, provide a more quantitative and application-oriented framework for UAV-borne GPR mission planning in cryosphere environments, which are largely absent from existing UAV-borne GPR reviews [25], [26]. Another important contribution of this manuscript is its explicit focus on cryosphere-specific operational and electromagnetic (EM) conditions that critically affect UAV-borne GPR performance. In particular, Section VI systematically incorporates environmental factors—including refractive index and dielectric constant variations caused by temperature gradients in ice, snow-density variability, and wind-induced UAV altitude perturbations—into the analysis of GPR signal propagation and data acquisition stability. Moreover, the manuscript critically examines the technical barriers limiting UAV-borne GPR deployment in cryosphere environments and discusses strategies for onboard processing, energy supply, and system optimization in large-scale cryospheric surveys in Section VI. This system-level perspective provides a more comprehensive understanding of the trade-offs governing UAV-borne GPR operation in harsh cryospheric conditions than prior UAV-GPR reviews that primarily focus on subsurface inversion in relatively stable urban or agricultural environments [25], [26].

## Advances in Cryosphere Subsurface Detection with Non-UAV GPR Technologies

GPR has been extensively applied for subsurface investigations of the cryosphere, offering quantitative information on ice thickness, internal layering, and basal conditions [23], as shown in Fig. 1. Prior to the recent emergence of UAV-borne systems, non-UAV GPR technologies, including ground-coupled GPR, fixed-wing airborne GPR, satellite-borne radar sounder, and helicopter-borne GPR, represented the main tools for large- and small-scale cryosphere surveys [25], [23]. Although these modalities share common electromagnetic sensing principles, their operational characteristics, achievable penetration depth, spatial resolution, coverage efficiency, and susceptibility to environmental perturbations differ substantially across cryosphere environments. In particular, the transferability of radar configurations and processing workflows is strongly governed by medium-dependent electromagnetic properties, including dielectric contrast, attenuation, scattering, and surface roughness. Consequently, system configurations optimized for shallow snowpack characterization are often unsuitable for deep glacier sounding or heterogeneous permafrost investigations. This section therefore not only summarizes representative non-UAV GPR technologies for cryosphere exploration, but also critically examines their methodological trade-offs, environmental adaptability, and practical limitations that motivate the development of UAV-borne GPR systems.

### Ground-coupled GPR

Ground-coupled GPR systems provide the most direct and high-resolution means of subsurface detection in cryosphere studies. With antennas in contact with the surface, they provide strong signal penetration and precise vertical resolution, ideal for measuring snow/ice thickness, mapping internal layers, and identifying englacial or subglacial hydrological features.

High-resolution surveys reveal spatial and temporal variations in glacier morphology, improving understanding of ice dynamics and climate-driven changes. For example, dense GPR profiles on Austre Lovénbreen, Svalbard, produced a 10 m grid bedrock digital elevation model, showing asymmetric bedrock, crevasses, supraglacial streams, and ice-volume distribution [27]. Similarly, temperate glaciers in Austria's Hohe Tauern were surveyed with 20 MHz GPR, bedrock topography and ice thickness were reconstructed via kriging and mechanical models, providing reliable ice-volume estimates [28]. Such surveys provide critical inputs for modeling glacier mass balance and subglacial water flow, highlighting the importance of precise ice-thickness mapping.

Other studies demonstrate ground-coupled GPR's utility in monitoring glacier change and hydrology. Pedersenbreen exhibited a polythermal structure with temperate ice in the lower cirque using 100 MHz and low-frequency radar, with 12% ice-volume loss over 73 years [29]. Ground-based surveys of Kongsfjorden tidewater glaciers (~1700 km) produced subglacial topography and ice thickness, revealing bedrock troughs influencing retreat potential [30], while 10–80 MHz GPR surveys at Leverett Glacier, Greenland, improved bed elevation and subglacial water routing estimates [31]. Multi-temporal GPR surveys on Ariebreen indicated a 73% ice-volume reduction between 1936 and 2007, with accelerated lower-elevation thinning [32]. Ground-coupled GPR also tracks snow accumulation and temporal variability. For instance, on Nordenskjöldbreen, accumulation in the period of 1986–1999 was detected 12% higher than that of 1963–1986 [33].

Advanced inversion algorithms, like glacier thickness estimation algorithm (GlaTE), optimize thickness estimates from sparse data [34], while high-density three dimensional (3D) GPR surveys reveal subglacial channels, englacial conduits, and drainage networks, validated with drone imagery and seismic data [35]. Recent Antarctic and Tibetan Plateau studies quantified polythermal structures, accumulation, and ice-thickness dynamics [36], [37], with techniques (e.g., stochastic dispersive modeling and particle swarm optimization-corrected based trimmed mean-constant false alarm rate) enhancing interface detection under low signal-to-noise ratios (SNR).

Overall, ground-coupled GPR remains the most reliable modality for high-resolution cryosphere characterization because direct antenna coupling minimizes transmission loss and maximizes signal penetration stability. These systems are particularly effective for shallow-to-intermediate depth investigations requiring fine stratigraphic resolution, such as snow accumulation mapping, englacial layer tracing, and subglacial hydrological characterization. However, their performance degrades in environments with severe surface roughness, extensive crevassing, or large-area coverage requirements, where logistical accessibility rather than radar capability becomes the dominant constraint. Non-contact GPR extend these observations to regional and continental scales by mounting antennas on aircraft, helicopter, or UAV, overcoming logistical constraints and enabling rapid data acquisition across vast, inaccessible glaciated regions [25], [26].

Moreover, although low-frequency antennas improve penetration in temperate or water-rich ice, their large physical dimensions and deployment complexity limit operational efficiency in remote polar environments. Many signal-processing techniques developed for ground-coupled GPR, including migration, clutter suppression, and layer-tracking algorithms, remain transferable to UAV-borne systems. In contrast, assumptions relying on stable antenna–surface coupling are generally not transferable because UAV-borne geometries introduce additional challenges including variable incidence angles, motion-induced phase instability, and reduced backscatter strength.

**Fixed-wing Airborne Radar Sounder**

An important category of non-contact GPR for cryosphere subsurface exploration is fixed-wing airborne GPR, or radar sounder. Originally designed for glaciology [38], these systems leverage the low attenuation of ice [23] to penetrate several kilometers, depending on frequency and temperature [39]. Early very-high-frequency (VHF) radar experiments in the 1950s–60s confirmed the feasibility of ice-thickness measurement [40], producing the first continuous profiles over Greenland and Antarctica [41], [7]. Subsequent efforts by the Scott Polar Research Institute [42] and the Technical University of Denmark [43] established systematic radar-sounding methods, including mapping ice-sheet topography using 60-MHz radar sounders [39].

Later, the University of Kansas developed the 150-MHz Improved Coherent Arctic Radar Depth Sounder (ICARDS) radar in the 1990s [44]. Subsequent systems incorporated multi-channel and array-processing capabilities [45]. These advances enabled 3D imaging of ice-sheet interiors and basal conditions in complex terrains [46]. Operating at a central frequency of 195 MHz with a 30-MHz bandwidth and variable peak power up to 1200 W, these radar sounders achieved more than 4 km penetration in field campaigns. Specialized systems were also developed for targeted depths: L-band (590–910 MHz) for shallow sounding (<500 m) [38], snow radars (center frequency of 5 GHz, bandwidth of up to 6-GHz) for layer retrievals (<40 m), and Ku-band for near-surface stratigraphy (<10 m) [38]. Low-frequency (1–30 MHz) radar sounders further improved performance over temperate glaciers [47], where signal attenuation is high. Comprehensive advancements in fixed-wings airborne radar sounders are detailed in [48].

International cooperation airborne radar surveys have been pivotal for mapping Antarctic and Greenland subglacial environments. Foundational efforts such as the SPRI–NSF–TUD campaigns (1960s–1970s) produced the first global ice-thickness compilations [49], later refined through Bedmap 1 [50], Bedmap 2 [51], Bedmap 3 [52], and ongoing SCAR Rings projects [53], now covering over 1.9 million line km. The BedMachine Antarctica dataset further improved bed-elevation estimates by integrating radar and mass-conservation models [54]. Collectively, these initiatives have greatly advanced understanding of ice-sheet geometry, basal conditions, and sea-level change.

Building upon extensive radar datasets, recent developments in artificial intelligence (AI) have automated and improved radar-data interpretation. Xu et al. [55] implemented deep learning in ice-penetrating radar with a 3D convnets and recurrent neural networks model for ice–air and ice–bed detection, where later works applied convolutional architectures. For instance, Rahnemoonfar et al. [56] used multi-scale CNNs for internal layer mapping, and Varshney et al. [57] developed fully convolutional networks for Greenland accumulation layers detection. Deep learning has also enhanced basal and hydrological analyses. For example, Wang et al. [58] used segmentation networks for basal classification, while Dong et al. [59] and Ma et al. [60] applied clustering and ensemble models to identify new subglacial lakes. Targeted AI applications now address depth-hoar mapping [61], low-SNR horizon tracing [62], and cross-domain modeling [63], all of which underscoring AI's growing role in cryosphere radar interpretation.

Fixed-wing airborne radar sounders achieve unparalleled large-scale coverage and kilometer-scale penetration because low-frequency operation substantially reduces electromagnetic attenuation in cold ice. However, this capability is fundamentally restricted by reduced spatial resolution, large antenna size, and high platform power requirements. Consequently, these systems are well suited for continental-scale ice-sheet mapping but are less effective for fine-scale near-surface investigations requiring low-altitude operation and high lateral resolution. Furthermore, many airborne radar configurations are optimized for relatively smooth and slowly varying ice-sheet geometries; their processing assumptions may become invalid in complex alpine or permafrost terrains where strong topographic variation and heterogeneous dielectric structures induce severe off-nadir clutter and migration ambiguity. Although advanced SAR focusing and AI-assisted interpretation frameworks developed for airborne radar sounding are highly transferable to UAV-borne GPR, the associated power consumption, antenna dimensions, and platform stability requirements remain major barriers for direct UAV implementation.

**Satellite-borne Radar Sounder**

The ability of GPR to probe ice has inspired satellite-borne radar systems for planetary and terrestrial exploration in super-large scale. Low-frequency sounders like SHARAD [64] and MARSIS [65] have orbited Mars to investigate subsurface structures, with MARSIS data providing indirect evidence of liquid water on Mars [66]. Similarly, JAXA's SELENE radar instruments have been used to probe the Moon's subsurface [67]. These achievements highlight the broader applicability of spaceborne radar sounders for investigating ice and subsoil in extraterrestrial context. On Earth, satellite observations reveal Arctic sea-ice loss and thinning at the margins of Greenland and Antarctic ice sheets [3], with multi-mission analyses showing accelerating ice-mass loss, highlighting satellites' key role in tracking cryosphere change and informing sea-level projections [19].

Recent advances have improved polar ice probing from orbit. Initial performance simulations show that reliable detection of internal horizons and basal interfaces requires high SNR, with grounded ice demanding over 85 dB for a 45-MHz radar with 10-MHz bandwidth [68]. The BingSat-TOPIS multistatic concept uses one high-power illuminator and multiple Mirror synthetic aperture radar (SAR) CubeSats to achieve stereoscopic radar sounding, compensating for up to 65 dB ice attenuation and detecting subsurface reflectors with reflectivity down to −3.5 dB, with tomographic resolution around 100 m × 20 m × 5 m [69]. A matched-filter back-projection algorithm processes multistatic CubeSat signals, accurately reconstructing volumetric subsurface profiles while maintaining phase coherence under moderate ionospheric disturbances [70]. Shallow-depth retrievals from CryoSat-2, enhanced with radar freeboard, penetration-rate corrections, and snow-depth inputs, improve sea-ice and near-surface ice-thickness estimates [71]. These advances extend airborne GPR principles to orbital platforms, enabling subsurface imaging from space. However, achieving tomographic resolution suitable for fine-scale investigations remains a significant challenge [69].

Despite their unmatched spatial coverage, satellite-borne radar sounders remain fundamentally constrained by orbital altitude, limited transmitted power, and coarse tomographic resolution. As a result, these systems are primarily effective for large-scale structural characterization rather than localized high-resolution imaging. The severe free-space propagation loss associated with orbital platforms imposes stringent SNR requirements and restricts the detectability of fine-scale subsurface structures, particularly in heterogeneous or strongly attenuating cryosphere environments. Moreover, although multistatic and tomographic configurations improve volumetric reconstruction capability, their spatial resolution remains insufficient for many engineering-scale cryosphere investigations. These limitations highlight an important sensing gap between continental-scale satellite observations and localized ground surveys, motivating the development of flexible intermediate-scale platforms such as UAV-borne GPR.

**Helicopter-borne GPR**

Helicopters are increasingly utilized in airborne geophysical surveys, offering advantages in rugged terrains, steep fjords, and remote regions accessible by vessel support. Despite their shorter range compared to fixed-wing aircraft, they can be deployed from icebreakers to perform coastal surveys and assist field operations. This combination enables access to previously unreachable zones, such as the grounding zones of fast-flowing, and heavily crevassed glaciers. Helicopter-borne GPR effectively resolves subsurface structures in glaciers and snowpacks, and other cryosphere settings. Early studies in 2000 showed high-frequency frequency modulated continuous wave (FMCW) and short-pulse radars with high-gain antennas could map englacial drainage networks while suppressing clutter [72].

In 2008, Grenzgletscher, a temperate Alpine glacier, was surveyed with a helicopter-borne gated stepped frequency continuous wave (SFCW) GPR (50–150 MHz dipole antennas) performing 20 frequency sweeps per second across 512 steps, enabling continuous high-resolution profiling over hazardous, crevassed terrain [73]. Data processing detail by Catapano et al., included global positioning system (GPS)-based height correction, time gating, frequency-domain transformation, and shifting zoom-based microwave tomography (MWT) inversion, producing tomographic images that identified crevasses, large rocks, and bedrock [24]. By 2010, helicopter-borne GPR applications expanded to environmental monitoring, such as detecting thin oil layers under snow with 1000 MHz GPR [74], and quasi-3D surveys over rock glaciers delineated shear zones and faulting patterns that are otherwise inaccessible [75]. Comparative studies highlighted that low-frequency pulsed antennas outperform high-frequency systems in water-rich ice [76]. Custom systems like AIR-ETH, combined with MATLAB processing, filter helicopter interference and enable reverse-time migration, recovering over 90% of ice-thickness profiles [77]. Dual-polarization antennas and advanced approaches further improve depth penetration and reflection coherence in complex terrains [78]. Overall, helicopter-borne GPR has been demonstrated to be able to provide efficient, high-resolution subsurface mapping in challenging cryosphere environments, with system design and processing strategies being crucial for data quality and interpretation.

Helicopter-borne GPR provides an important intermediate solution between fixed-wing airborne systems and ground-based surveys by combining relatively low-altitude operation with improved regional accessibility. This geometry enables higher-resolution imaging in steep or hazardous terrains where fixed-wing aircraft are operationally constrained. However, rotor-induced vibration, platform electromagnetic interference, and limited trajectory stability introduce significant challenges for coherent radar imaging and motion compensation. Although many helicopter-borne processing workflows—including SAR focusing, tomographic inversion, and clutter suppression—are directly transferable to UAV-borne GPR systems, the substantially lower payload capacity and power availability of UAV platforms impose stricter constraints on antenna configuration, onboard processing, and survey endurance. Consequently, helicopter-borne systems demonstrate the feasibility of airborne high-resolution cryosphere radar imaging while simultaneously revealing the key engineering challenges that UAV-borne implementations must overcome.

**Summary of Non-UAV GPR Technologies in Cryosphere**

Collectively, existing non-UAV GPR technologies demonstrate that no single radar configuration is universally optimal across all cryosphere environments. Instead, system performance is fundamentally governed by trade-offs among penetration depth, spatial resolution, platform mobility, power consumption, and environmental adaptability. Low-frequency radar systems are advantageous for deep glacier and ice-sheet sounding because of reduced attenuation, but they require large antennas and high transmitted power that limit deployment flexibility. Conversely, higher-frequency configurations provide superior near-surface resolution for snowpack and shallow ice investigations, yet suffer from limited penetration depth in water-rich or heterogeneous media. Similarly, signal-processing workflows developed for stable airborne or ground-coupled geometries are only partially transferable to UAV-borne systems because low-altitude operation introduces stronger motion perturbation, variable antenna–surface coupling, and stricter payload constraints. These limitations collectively motivate the emergence of UAV-borne GPR as a complementary sensing modality capable of bridging the operational gap between localized ground surveys and large-scale airborne observations.

## An Overview of UAV-Borne GPR Technology

Although conventional non-contact GPR platforms, i.e., fixed-wing aircraft, satellites, and helicopters, offer broad spatial coverage, they remain costly, environmentally demanding, and limited in resolution due to the large antenna-to-surface distance. In addition, flight-line spacing cannot approach the dominant GPR wavelength, preventing high-density data acquisition. Recently, UAV-borne GPR has emerged as a promising alternative, enabling flexible, high-resolution, and cost-effective surveys. Rapid progress in autonomous navigation, lightweight radar design, multisensor integration, and efficient power systems has driven this advancement. With continued miniaturization of radar subsystems and the typically low-interference environment of cryosphere regions, GPR is now increasingly integrated into small-to medium-scale UAVs for detailed shallow cryosphere investigations.

### Extensive Utilization of UAV-borne GPR Systems

Recent research on non-contact GPR increasingly focuses on drone- and UAV-borne systems. This trend is driven by advances in autonomous flight control, multi-sensor integration, avionic systems, energy-efficient batteries, and reduced operational costs [79]. Early demonstrations of UAV-borne radar imaging were reported in [80], [17], followed by initial experiments using high-frequency radars [81]. Although limited in penetration, these systems confirmed the feasibility of UAV-borne imaging radars. A subsequent feasibility analysis of multifrequency GPR for rotary UAVs [82] simulated operations at 500 MHz and 4 GHz to evaluate target responses of varying dimensions. Later, an FMCW GPR employing two log-periodic dipole array antennas on a mini-UAV was developed for archaeological surveys [83]. Operating at 745 MHz with 510 MHz bandwidth, it successfully detected a buried object at 0.4 m depth. Similarly, a lightweight software-defined GPR for landmine detection [84] demonstrated real-time configurability and reliable detection up to 0.2 m in moist soil. Further progress included a compact radar prototype integrated with an MWT-based processing framework [85], using a 3.95 GHz pulsed radar (1.7 GHz bandwidth) to validate cost-effective UAV radar imaging. A wideband bistatic FMCW GPR (1–4 GHz) [86] optimized for weight, power, and penetration was later flight-tested, confirming effective detection of corner reflectors.

Demonstrating the potential of UAV-borne GPR for precision agriculture and environmental monitoring, a lightweight, drone-mounted frequency-domain GPR successfully mapped the top 10–20 cm of soil with full-waveform inversion retrieving both permittivity and corresponding moisture distributions [87]. Parallel efforts focused on polarimetric side-looking SAR on UAVs demonstrated the detection and classification of small metallic objects, including anti-personnel mines, by exploiting the polarization dependence of radar cross-sections [88]. Subsequent work improved subsurface focusing by integrating InSAR-derived digital elevation models (DEMs) with UAV-borne ground-penetrating SAR (GPSAR), enabling accurate detection of shallowly buried objects on non-planar surfaces and enhancing the SNR of buried targets [89]. Fernandez et al. further advanced UAV-borne GPR systems by addressing uncertainties in airborne GPR-SAR imaging, correcting height-induced distortions, UAV tilting effects, and refining singular value decomposition (SVD) filtering for clutter suppression, achieving up to 100% detection of metallic and non-metallic targets under controlled scenarios [90]. These improvements were complemented by the development of multichannel array-based UAV-GPR systems, which increased scanning throughput by up to four times while maintaining centimeter-level resolution, improving clutter suppression, and enabling reliable detection of small or challenging targets [91].

Later, the integration of complementary sensing modalities was explored. A 2023 joint UAV system combined GPR and magnetometer sensors for landmine detection, leveraging finite-difference time-domain (FDTD)-based GPR modeling with SVD and Kirchhoff migration for precise hyperbola focusing and MAG-based magnetic anomaly estimation [92]. Recent advances apply amplitude inversion algorithms to drone-based GPR, iteratively reconstructing subsurface structures and permittivity. Using the known air layer at UAV altitude, these methods reduce input uncertainty, enabling robust, frequency- and path-independent subsurface characterization [93]. Additional studies on UAV-borne GPR are summarized in [25] and [26]. Further details on UAV-borne GPR applications in the cryosphere are provided in Section IV.

### Local Positioning and Georeferencing for UAV-borne GPR

Accurate positioning and georeferencing are essential for UAV-borne GPR systems, as they determine both imaging resolution and the smallest detectable target size. When georeferencing errors fall below the GPR wavelength, SAR techniques can be applied to improve image focusing and cross-range resolution beyond the intrinsic antenna limits [94].

Typical UAVs employ onboard sensors such as inertial measurement unit (IMU), barometer, compass, and global navigation satellite system (GNSS) receiver, providing positional accuracies from several meters to a few decimeters. For centimeter-level precision, advanced techniques like GNSS real-time kinematics (RTK) [95] are required, offering real-time positioning with significantly higher accuracy. However, vertical positioning remains a major challenge since GNSS altitude refers to an ellipsoidal or geoidal model rather than the true terrain. To address this, laser rangefinders and radar altimeters are often integrated, providing direct surface measurements. Comparative studies among GNSS-RTK, laser rangefinding, and radar altimetry have shown that direct ranging sensors can substantially improve ground profile estimation accuracy [96].

Leveraging these advances, several high-precision UAV radar imaging systems have been developed. The University of Oviedo's platform [97] employed a DJI Matrice 600 Pro with a carrier-phase differential (CD) GPS-RTK module and a P440 pulsed radar (3.1–5.1 GHz), using circularly polarized helix antennas to balance penetration depth and resolution. Similarly, the German Aerospace Center designed a UAV radar imaging system [98] combining RTK-GPS and IMU for real-time trajectory reconstruction, equipped with an FMCW radar (500 MHz–3 GHz) supporting both slanted SAR and down-looking GPR modes.

Further refinements were reported in [99], where an upgraded version of the system in [85] used CDGPS-based motion compensation to correct altitude variations and enhance focusing accuracy. Continued progress in GNSS-RTK technology [100], complemented by precise point positioning (PPP), has further improved UAV trajectory estimation and data georeferencing. Despite PPP's limited real-time applicability due to long convergence times, the adoption of dual- and triple-frequency RTK modules [101] has markedly enhanced performance while reducing system costs.

### UAV-Borne GPR Working Principle

To illustrate the basic working principle of UAV-borne GPR, a simplified two-dimensional geometry, as shown in Fig. 2, is normally considered. The scene comprises a two-layer medium in which the upper half-space is air and the lower half-space is ice, representing a typical configuration for airborne sounding over ice sheets [102].

At a fixed UAV height $Z_0$ above the air–ice interface, the range-dependent amplitude of the received signal is recorded as an A-scan [23]. In practical environments containing multiple reflectors, the overlap of echoes complicates the interpretation of individual A-scans. Therefore, consecutive A-scans acquired along a flight line shown in Fig. 2, are combined to form a B-scan, which represents a vertical cross-section of the subsurface beneath the flight path.

For the simplified geometry shown in Fig. 2, a point-like target located at depth $z$ would ideally appear as a single point scatter in the B-scan. In practice, due to the finite antenna aperture and corresponding beamwidth, a UAV-borne GPR positioned at a horizontal offset $x_d$ from the target also receives scattered signals, causing the reflected echo to appear as a hyperbolic trajectory in the B-scan. During propagation, the EM wave experiences absorption, scattering, refraction, and a reduction in velocity [23]. Refraction follows Snell's law, with the refraction point horizontally displaced by $x_b$ from the target. The incident and refracted angles are denoted by $\theta_0$ and $\theta_1$, and the corresponding propagation paths above and under the air–ice interface are $R$ and $r$, respectively.

The propagation characteristics of GPR signals depend on the medium's relative permittivity, electrical conductivity, and wave velocity. Signal attenuation and scattering losses increase with frequency, which constrains the effective penetration depth. Consequently, UAV-borne GPR systems typically operate within the sub-GHz to lower-GHz range [103]. Because the permittivity of natural materials is often uncertain, accurate depth estimation of subsurface targets remains challenging; in many cases, only the lateral extent or geometry of buried structures can be reliably inferred.

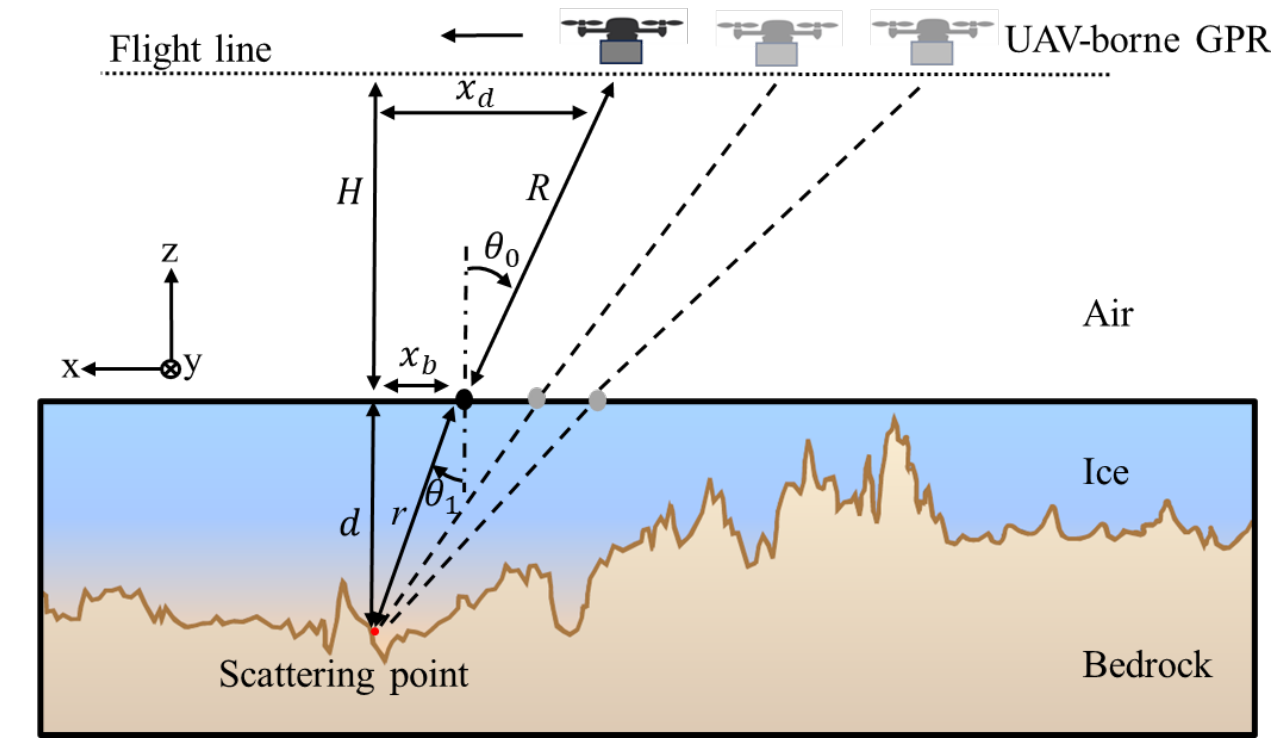

**Fig. 2.** Illustration of EM wave propagation in a UAV-borne GPR system for sub-ice detection.

When multiple B-scans are collected along parallel flight lines, they can be integrated to generate a C-scan [104], providing volumetric imaging of the subsurface, which effectively refers to the 3D reconstruction. The horizontal resolution of the 3D image is determined by both the antenna beamwidth and the spacing between adjacent flight lines. Owing to the constraints of payload capacity and power on UAV platforms, achieving high-resolution subsurface imaging typically requires additional signal processing and reconstruction techniques, which are discussed in the following subsection.

### Data Processing

#### *Pre-processing techniques*

UAV-borne GPR datasets require systematic preprocessing to ensure spatial, temporal, and amplitude consistency prior to interpretation [26], [105]. The workflow begins with spatial interpolation at uniform intervals using GNSS coordinates to correct for variations in UAV flight speed and altitude [106], which constitutes a fundamental step of the procedure. To further enhance uniformity, profiles may be additionally resampled or spatially flipped to align flight directions and compensate for non-optimal time-window settings [105]. In some low-frequency datasets, direct current (DC) offset removal is occasionally required to correct baseline shifts [107]. A time-zero correction is an essential step that aligns all traces with the transmitter firing instant, typically determined through system calibration and amplitude-threshold detection of the direct air wave [108]. Background or mean-trace removal is routinely performed to suppress the emitted pulse and antenna ringing, either across the full profile or within a sliding window [109]. To improve signal clarity, a band-pass Butterworth filter is routinely employed to suppress residual high- and low-frequency noise [110], and a spatial median filter would be beneficial for mitigating constructive ringing [105], [106]. Static or topographic correction is indispensable for compensating UAV altitude variations, achieved by either time-shifting traces according to two-way travel time or applying Fourier phase-shift adjustments to maintain a smooth acquisition surface [26], [105].

Furthermore, relative alignment of adjacent flight lines (inter-line correlation adjustment) is often implemented to correct small lateral offsets between alternating flight lines caused by timing delays [35]. Amplitude correction forms a key part of the workflow, compensating for geometrical spreading and attenuation through power-law or time-varying gain functions depending on materials properties of subsurface, e.g., snow wetness [111]. For enhanced reflection sharpness, deconvolution may be optionally introduced to remove source-wavelet effects [112], and temporal interpolation via Fourier transform can be advantageously applied to increase trace density for improved data visualization and quantitative analysis [104].

***Diffraction and reflection wavefield separation***

GPR primarily observes two wavefield phenomena: diffractions and reflections. Diffractions originate from scattered EM waves caused by subsurface anomalies, such as landmines [84], tree roots [113], embedded rebars [79]. These signals typically appear as hyperbolic patterns in radargrams, contrasting with the continuous wavefronts produced by reflections from planar interfaces like the air–ground boundary, layered soils and ice, or the ice–bedrock interface [23].

Diffraction wavefields play a crucial role in GPR applications. First, researchers can infer the position, size, and shape of subsurface anomalies from their diffraction characteristics. Second, diffraction-focusing-based techniques have been widely recognized as effective methods for modeling the velocity structure of subsurface media [114], thereby facilitating subsequent processes such as subsurface imaging. However, diffracted wavefields are often difficult to isolate due to aliasing with strong reflected wavefields, which typically exhibit much higher amplitudes than those generated by local scatterers. This challenge is particularly pronounced in UAV-borne GPR, where strong scattering often originates from the air–ground interface, while signals from subsurface anomalies are generally weak due to high EM attenuation. Consequently, wavefield separation becomes a critical step in UAV-borne GPR data processing.

One example of a wavefield separation technique specifically designed for UAV-borne GPR is the common-reflection-surface (CRS)-based coherent stacking and subtraction scheme [115], adapted from CRS stacking in seismic processing [116]. First, the radar echo data $G$ is modeled as

$$G(x_0,t_0) \approx C_{\mathrm{ref}}(x_0,t_0) + C_{\mathrm{diff}}(x_0,t_0) + \mathcal{N}(x_0,t_0), \quad (1)$$

where $C_{\mathrm{ref}}$ and $C_{\mathrm{diff}}$ are the coherent reflected and diffracted wavefields, respectively, and $\mathcal{N}(x_0,t_0)$ is random noise. Reflection separation is achieved via coherent stacking over a local aperture n centered at the data point $(x_0,t_0)$ :

$$C_{\mathrm{k}}(x_0,t_0) \approx \frac{1}{\mathrm{n}}\sum_{i=1}^{\mathrm{n}} G[x_{\mathrm{m}},t_{\mathrm{k}}], \quad (2)$$

where n is the aperture size, $x_{\mathrm{m}} = x_0 + \Delta x_{\mathrm{m}}$, with m indexing the single-channel A-scan, and $\Delta x_{\mathrm{m}}$ the spatial offset, $t_{\mathrm{k}}$ denotes to the directional filtering function.

The core of the CRS-based coherent stacking and subtraction scheme is as follows. First, to improve diffraction–reflection separation, a directional filtering function $t_{\mathrm{k}}$ is designed to better match the reflected wave, expressed as

$$t_{\mathrm{k}} = t_{\mathrm{CRS-GPR}}(\Delta x_{\mathrm{m}}) = t_0 + \frac{2\,sin\,\alpha}{v_0}\Delta x_{\mathrm{m}}, \quad (3)$$

where $v_0$ denotes the near-surface EM wave velocity [34]. The optimal $sin\,\alpha$ is determined by maximizing the following objective function:

$$\max_{sin\alpha} \frac{1}{\mathrm{n}} \frac{\sum_{\delta t}\left\{\sum_{\mathrm{m}=1}^{\mathrm{n}} G\left[x_0+\Delta x_{\mathrm{m}}, t_0+\frac{2\sin\alpha}{v_0}\Delta x_{\mathrm{m}}\right]\right\}^2}{\sum_{\delta t}\sum_{\mathrm{m}=1}^{\mathrm{n}} G^2\left[x_0+\Delta x_{\mathrm{m}}, t_0+\frac{2\sin\alpha}{v_0}\Delta x_{\mathrm{m}}\right]}. \quad (4)$$

where $n$ and the time window $\delta_t$ govern the stability of diffraction separation. However, a wide range requires intensive computational resources. Therefore, selecting an optimal range for these parameters is beneficial for the entire optimization process. The optimization procedure of these two parameters is summarized as follows [115]. The search interval for $n$ is determined automatically based on the radius of the first Fresnel zone ($r_f$) [117]. In seismic studies, the dominant energy of a reflection event is primarily concentrated within this Fresnel zone at the observation surface [118], [119]. Following this, when applying GPR processing in the spatial domain, offset between a trace and $x_0$ is constrained

$$\Delta x_m \leq r_f = \frac{1}{2}\sqrt{\lambda T} \quad (5)$$

where $\lambda$ represents the wavelength of the EM wave, and T denotes the distance between the GPR transmitter Tx and the reflecting interface. Based on this relationship, the search limit for $n$ can be expressed as

$$n \leq \frac{\frac{1}{2}\sqrt{\lambda v_0 t_0}}{\Delta d} \quad (6)$$

where $\Delta d$ is the spatial spacing between two adjacent GPR A-scans. (6) indicates that the allowable range of $n$ depends on the location of the data point rather than remaining fixed. Such adaptivity is consistent with the characteristics of GPR imaging, where the resolution decreases with increasing depths and therefore requires a larger aperture size $n$. In addition, the parameter $\delta_t$ is assigned the same value as $n$, this unified setting has been demonstrated to be efficient [115], [120]. Once the reflected wavefield is obtained from (2), the desired diffracted components can be extracted by subtracting the reflected signals from the composite data described in (1).

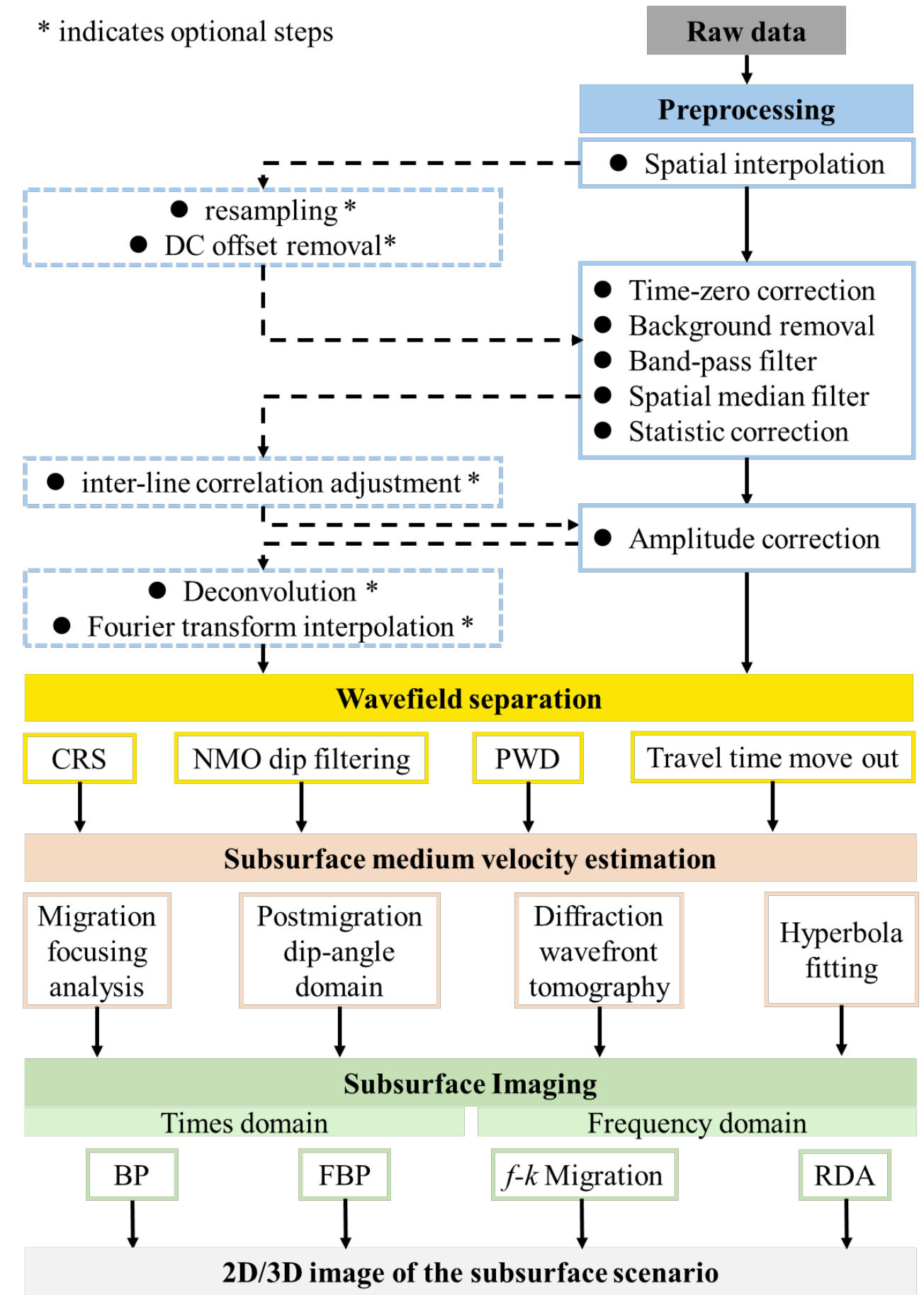

**Fig. 3**. The processing flowchart for UAV-borne GPR data.

Aside from CRS-based coherent stacking and subtraction scheme, other GPR data wavefield separation techniques that could be suitable for UAV-borne GPR systems are:

- **The normal move out (NMO) dip filtering:** This method applies normal moveout correction to flatten horizontal reflectors and suppress dipping events, enabling separation of reflections with different dips [121].
- **Planewave destruction (PWD):** This technique estimates the local slope of events and predicts plane waves accordingly, allowing selective removal of coherent noise or separation of overlapping signals with complex dips [122].
- **Travel time move out/horizontal slowness (HS) based coherent staking:** This method generate a dataset dominated by reflections, which can then be subtracted from the original data to isolate diffractions. It represents a direct application of the CRS stacking technique from seismic data processing, to ground-coupled GPR data [123].

The efficiency of the above methods, when applied to UAV-borne GPR, is compared using both simulation and field data as Fig. 4, 5, 11-13 in [115].

***Subsurface medium velocity estimation***

Accurate estimation of subsurface velocity is a fundamental step in GPR reflection imaging, as it governs the precision of imaging and depth conversion processes. In conventional GPR surveys, particularly those acquired with a single bistatic antenna pair at a fixed transmitter–receiver offset, offset-dependent information is generally unavailable, which is a limitation commonly encountered in UAV-borne GPR. Under such conditions, diffraction-based velocity analysis becomes especially advantageous, as it enables reliable velocity estimation directly from diffraction signatures without requiring multi-offset data [124].

Diffraction-based methods for velocity estimation can generally be categorized into four major approaches:

- **Migration focusing analysis:** This approach is based on the principle that a diffraction event is focused to a single point when imaged with the correct velocity. By evaluating the level of focusing across a range of migration velocities, the subsurface velocity field can be inferred [125]. The sharpness of focusing can be quantified using techniques such as velocity continuation, local kurtosis [126], and local diffraction semblance [127]. This method has also been extended to 3-D azimuthally anisotropic velocity analysis [128].
- **Postmigration dip-angle domain analysis:** In this strategy, velocity estimation is performed in the dip-angle common image gather domain. In this representation, specular reflections appear as upward-curving hyperbolae aligned with the reflector dip, whereas diffractions become flat only when the correct migration velocity is applied. The velocity is then determined by measuring the flatness of diffraction events [129].
- **Diffraction wavefront tomography:** This approach utilizes tomographic inversion of diffraction wavefront attributes. It is based on the concept that the two-way travel time and local wavefront parameters can approximate the kinematic behavior of reflection or diffraction events [130]. This method has demonstrated robustness in strongly heterogeneous media.
- **Hyperbola fitting analysis:** This approach utilizes the geometric relationship between the two-way travel time and antenna offset of diffraction events generated by point-like reflectors. The reflected wavefront forms a hyperbolic trajectory on the radargram [23]. By iteratively matching a theoretical hyperbola to the observed diffraction curve, the velocity that provides the best fit can be estimated [131]. In general, for a target at a given depth, a broader (flatter) hyperbola corresponds to a higher propagation velocity, whereas a steeper curvature indicates a lower velocity.

Among these, migration focusing analysis is the most widely applied approach for diffraction-based velocity analysis in GPR data [132]. A recently proposed weighted negative entropy (WNE)–based method, derived from the trans-medium propagation model of air-coupled EM waves, exemplifies this approach and is specifically designed for UAV-borne GPR [115]. In this method, negative entropy (NE) is employed to evaluate the local focusing level due to its statistical robustness and superior sensitivity compared with local semblance. Additionally, a weighting strategy is introduced to suppress noise effects. First, for the diffraction-only B-scan data, each pixel $(t_0, x_0)$ is assigned a weighting factor

$$W(t_0, x_0) = 1 - 0.5 + 0.5\left[\frac{M(t_0, x_0)}{M_{max}}\right], \quad (7)$$

where $M(t_0, x_0)$ represents the noise pixel filter,

$$M(t_0, x_0) = \begin{cases} \sigma(t_0, x_0), c(t_0, x_0) > c_{threshold} \\ 0, c(t_0, x_0) < c_{threshold} \end{cases}, \quad (8)$$

and $c(t_0, x_0)$ denotes the maximum linear coherence of pixel $D(t_0, x_0)$ within the spatiotemporal window $(\delta t, n)$:

$$c(t_0, x_0) = \max_p \frac{1}{n} \frac{\sum_{\delta t}\left[\sum_{i=1}^{n} D(t_0 + p\Delta x_i, x_0 + \Delta x_i)\right]^2}{\sum_{\delta t}\sum_{i=1}^{n} D^2\left(t_0 + p\Delta x_i, x_0 + \Delta x_i\right)}, \quad (9)$$

where $p$ is the optimal slope corresponding to the maximum coherence $c(t_0, x_0)$, $n$ is the A-scan window size, and $\delta t$ is the time-window length. The filter amplitude $\sigma(t_0, x_0)$ in (8) is

$$\sigma(t_0, x_0) = \begin{cases} 0, t_0 \ in \ t_{air} \\ -\frac{1}{4}p^2 t\beta^2 v\left[1 - \left(\frac{1-\beta}{\beta}\right)^2\left(\frac{{v_c}^4}{v^4}\right)\right], t_0 \ in \ t_{med} \end{cases}, \quad (10)$$

where $v_c$ is the EM velocity in air, $t_{air}$ and $t_{med}$ are the propagation times in air and the subsurface, respectively, and $\beta = t_{med}/(t_{air} + t_{med})$ . Detailed interpretations of $\sigma$ , $c(t_0, x_0)$, and $\sigma(t_0, x_0)$ can be found in [115].

Based on these definitions, the only undetermined parameter in $W(t_0, x_0)$ is the medium velocity $v$, which serves as the primary estimation objective of this analysis. Combining $W(t_0, x_0)$ with the NE formulation yields the WNE velocity analysis method:

$$S_W(t, x, v) = F_v[D(t, x)W(t, x)]g_W(x, t) \times \log\left(F_v[D(t, x)W(t, x)]g_W(x, t)\right), \quad (11)$$

where $F_v$ denotes the any migration focusing operator applied with velocity $v$, which will be detailed in the subsection *iv*, and

$$g_W(x, t) = \frac{1}{\sqrt{\frac{1}{N}\sum_{i=1}^{N}[D_i(t, x)W(t, x)]^2}}. \quad (12)$$

By scanning over a velocity range $v \in [v_1, v_N]$, $N$ focused B-scan images can be generated, forming a 3D data cube ($t$, $x$, $v$). Each A-scan corresponds to a $t$-$v$ plane, from which the velocity $v$ corresponding to the maximum amplitude at each time $t$ is extracted to construct the 2D velocity distribution along the survey line.

***Subsurface imaging***

Imaging of subsurface targets using GPR technology employs advanced focusing algorithms [24], [26], which serve as critical components of GPR data processing. Commonly referred to as SAR processing, these algorithms are designed to migrate subsurface point targets—typically appearing as diffraction hyperbolas—into well-focused bright points [23]. The application of such algorithms markedly enhances the detection accuracy of subsurface interfaces and ice layers, yielding higher SNR and improved imaging resolution [104].

Based on the operational domain and intended imaging configurations, SAR techniques can be broadly categorized into time-domain and frequency-domain methods. The time-domain class primarily includes the back-projection (BP) algorithm [133] and its fast variants, such as fast BP (FBP) [134]. BP-based approaches offer high imaging accuracy and inherent adaptability to arbitrary trajectories and non-uniform sampling, making them suitable for UAV-borne GPR applications. Their primary limitation, however, is the high computational cost, which restricts real-time and large-scale processing.

In contrast, frequency-domain methods, represented by frequency–wavenumber (*f–k*) migration [135] and the Range–Doppler algorithm (RDA) [136], are implemented based on Fast Fourier Transform (FFT). These algorithms offer high computational efficiency and are well suited for linear aperture configurations, making them particularly appropriate for rapid processing over large coverage areas. Hence, selecting an appropriate SAR algorithm for UAV-borne GPR requires balancing imaging accuracy and computational efficiency under practical survey conditions.

One of the most widely used real-time imaging techniques is *f–k* migration, which offers notable flexibility for layered media by directly specifying the wave propagation velocities in layers instead of employing ray tracing in a predefined geometric model. Originating from wavefront reconstruction theory, *f–k* migration was initially developed for seismic signal processing. Its core principle involves downward back-propagating the recorded wavefield $s(x, 0, t)$ to its points of origin $(x, z, 0)$ within the subsurface, effectively collapsing diffraction hyperbolas into focused points. For UAV-GPR surveys in layered media, the implementation typically follows three main steps:

*Step-1:* Transform the received time-domain echoes $s(x, 0, t)$ into the frequency ($f$)-wavenumber ($k_x$) domain $S(k_x, 0, f)$ by applying a 2D FFT over the horizontal distance $x$ and time $t$ axis.

$$S(k_x, 0, f) = \int\int s(x, 0, t) \cdot \exp(-j2\pi ft - jk_x x)\, dxdt; \quad (13)$$

*Step-2:* Apply the following propagation equation:

$$S(k_x, z, f) = S(k_x, 0, f)\exp\left(jk_{z_{air}}h_0\right)\exp(jk_{zice}d), \quad (14)$$

where $S(k_x, 0, f)$ is first migrated from the UAV-GPR height $h_0$ to the surface level, and further migrated to depth $d$. Then integrating $S(k_x, z, f)$ over $f$ to reconstruct the time-domain field at specific time 0, which offers instantaneous amplitude at zero time delay.

$$S(k_x, z, 0) = \int_{-\infty}^{+\infty} S(k_x, z, f)\exp(j2\pi f \cdot 0)df = \int_{-\infty}^{+\infty} S(k_x, z, f)\, df; \quad (15)$$

*Step-3:* By performing inverse FFT over $k_x$,

$$s(x, z, 0) = \int_{-\infty}^{+\infty} S(k_x, z, 0)\exp(jk_x x)\, dk_x, \quad (16)$$

the final migrated image $s(x, z, 0)$ in physical space is yield.

A brief overview of other commonly adopted SAR algorithms is as follows:

- **RDA:** In the RDA, the two-way travel time between the radar antenna and subsurface point targets is first computed using an ideal geometric model that accounts for ray bending across layered interfaces. The received echoes are then transformed into the range–Doppler domain, where Range Cell Migration Correction (RCMC) compensates for range curvature by realigning target echoes into a single bin. Finally, azimuth compression is achieved by matched filtering the RCMC-corrected signal with a reference function derived from the modeled delays, thereby focusing the energy and attaining fine azimuth resolution [136].
- **BP:** For an arbitrary trajectory, the imaging plane is defined by the known GPR sampling positions and the geometric model. The BP algorithm iteratively reconstructs each pixel in the output grid. For a given pixel, the two-way travel time from each antenna phase center to the pixel location is computed and used to extract the corresponding echo from the radargram. Phase compensation is then applied, and the pixel intensity is obtained through coherent summation of all phase-corrected echoes [133].
- **FBP:** To mitigate the heavy computational load of conventional BP, fast BP methods—such as the Cartesian factorized back-projection—have been developed to accelerate processing by compressing the sub-image spectrum and reducing the required spatial wavenumber bandwidth. The algorithm employs a two-step spectrum compression procedure involving center alignment and inclination correction. Through this compression, the sub-image can be reconstructed alias-free at a lower Nyquist sampling rate, thereby significantly reducing computational complexity [134].

## An Overview of UAV-Borne GPR in Cryosphere Detection

Recent progress in UAV-borne GPR has extended its application from engineering and archaeological surveys to cryosphere studies, offering new possibilities for mapping shallow subsurface structures in frozen terrains. Building upon the technological foundations summarized in the previous section, this part focuses on the implementation of UAV-borne GPR in detecting and characterizing cryosphere shallow features. Table 1 summarizes the main characteristics of UAV-borne GPR systems reported in the literature, organized into three parts aligned with the following subsections—snowpack, ice forms, and permafrost—where studies in each part are arranged in order of increasing penetration depth.

### Snowpack Analysis

One of the most prominent applications of UAV-borne GPR in the international cryosphere research community is snow cover investigation [137]. In 2016, a study on UAV-mounted radar, constrained by size/weight, and the trade-off between frequency-dependent resolution and attenuation, conducted experiments on McMurdo Sound sea ice and identified an operational frequency range of 1.5–4.5 GHz as an optimal compromise [103]. Building on this, a custom UWB radar system (0.95–6 GHz) deployed on an octocopter drone with a wingspan of 1.5 m successfully demonstrated its capability in 2018 by detecting a target buried 1.5 m beneath wet snow at a flight altitude of 1 m [138], and was later used to profile dry snow depth on Arctic sea ice [139].

In [140], researchers developed a drone-mounted ultra-wideband snow sounder, addressing key challenges in antenna design (lightweight and compact form factor) and continuous-wave radar operation (range ambiguities and antenna isolation). Through novel processing techniques, the system accurately estimated total snow volume and snow water equivalent during a campaign in Svalbard. Furthermore, a lightweight UAV-mounted radar designed for snow depth measurements on sea ice was validated in Antarctic trials, demonstrating effective performance at varying speeds and altitudes, with accuracies ranging from ±3.2 cm (stationary) to ±9.1 cm (in flight), thus supporting large-scale survey applications [141].

With advancements in both hardware and software, UAV-mounted GPR has recently been applied not only to snow depth and area mapping [142], but also to snow water resource assessment [143]. For instance, one study demonstrated a multi-method system combining drone-based GPR, photogrammetry, and TDR to monitor snowpack properties. Using a 1.5 GHz GPR, the approach successfully tracked spatiotemporal variations and hydrological responses to rain-on-snow events, showing strong potential for broader hydrological applications [143].

Most recently, researchers have demonstrated the potential of UAV-borne GPR for snowpack characterization. Importantly, they have also identified the limitations of this specific GPR application in survey scenarios and established operational guidelines—including flight speed, altitude, and frequency bandwidth—based on extensive experiments [105]. In the context of snowpack detection, UAV-borne GPR offers a promising solution for the localization and rescue of avalanche victims [144]. Despite advances in safety measures, the annual number of avalanche-related fatalities has remained relatively stable over the past several decades [145]. Since the survival probability of a fully buried avalanche victim declines sharply within the first 15 minutes [146], rapid, wide-area, and multi-flight surveys are essential to ensure adequate coverage of extensive avalanche zones. However, the rapid deployment of a large fleet of manned aircraft within such a limited timeframe is logistically challenging. In contrast, UAVs, owing to their compact size and capacity for simultaneous multi-flight operations, are better suited to meet these demands. UAVs equipped with specialized devices have been employed to detect electronic equipment buried beneath snow, such as transponders or mobile phones, thereby enabling the indirect tracking of victims [147]. However, these approaches do not directly detect the victims themselves, which can lead to positioning inaccuracies and related limitations.

Pulse GPR systems operating in the sub-GHz range have been demonstrated to be effective for victim detection, as the reflection coefficient between human muscle tissue ($\varepsilon_{muscle}$=55.03) and snow ($\varepsilon_{snow}$=1.81) is sufficiently high [148]. Moreover, GPR can distinguish the radar cross-section of the human body from that of the surrounding snowpack [149]. In 2002, GPR systems operating at 900 MHz and highly mounted on sledges were tested in the Blåskumbreen study area in 2000 and in the Svalbard archipelago in 2001 [148] [150]. In these trials, researchers successfully identified victims buried at depths of up to approximately 4 m. These studies demonstrated promising results; however, the requirement for real-time image interpretation and the presence of complex avalanche debris, which generates significant clutter, have hindered the practical application of GPR [151]. This highlights the need to develop UAV–GPR systems equipped with intelligent software capable of indicating the probable locations of victims, thereby enabling effective use by non-technical rescue personnel.

A two-step approach, consisting of a snowpack extraction method and a diffraction enhancement algorithm, was proposed for airborne GPR [152]. This method was validated in experiments where the GPR was elevated 6 m and 12 m above the surface, corresponding to typical UAV operating heights. In addition, a data post-processing technique based on the statistical evaluation of reflected radar energy was developed, which successfully localized simulated victims within realistic ranges under avalanche conditions [153]. More recently, a UAV equipped with a side-looking FMCW radar operating from 1 to 4 GHz and a circular aperture was investigated, where a ground-penetrating synthetic aperture radar processor was implemented to successfully detect avalanche victim equivalents buried at depths of up to 40 cm [154].

Collectively, these studies demonstrate that UAV-borne GPR is particularly effective for snowpack investigations when operated in the high-MHz to GHz frequency range (typically ≥900 MHz to 6 GHz), where the relatively low attenuation and weak dielectric heterogeneity of dry snow enable high-resolution imaging of shallow stratigraphy. Ultra-wideband systems provide centimeter-scale vertical resolution and accurate snow-depth retrievals, making them suitable for snow-water-equivalent estimation, avalanche rescue, and near-surface hydrological monitoring. However, the same frequency configurations exhibit limited penetration capability in wet snow or deeper cryosphere targets due to increased dielectric loss and scattering. Consequently, snow-oriented UAV-GPR systems generally prioritize bandwidth, compact antenna geometry, and rapid areal coverage over deep penetration performance. In practical deployment, low-to-medium-altitude flight (<0.5–2 m) and relatively high radar pulse repetition frequency (PRF) are often required to maintain sufficient SNR and coherent imaging quality, while lightweight antenna configurations remain critical for multirotor UAV endurance.

From a system-design perspective, UAV-GPR configurations optimized for snowpack analysis are not directly transferable to glacier or permafrost investigations. Frequencies above 1 GHz provide high resolution but lose penetration capability in heterogeneous or water-rich frozen media. Moreover, compact UWB antennas become increasingly inefficient at lower frequencies required for deeper sounding. Therefore, snow-focused systems are best suited for shallow targets (within several meters) requiring fine-scale stratigraphic characterization rather than deep subsurface imaging.

**Ice Forms Detection**

Beyond snowpack, the application of radar in cryosphere science extends to determining the internal structure and thickness of ice—from glaciers and lake ice to ice sheets [23], [155]—using frequencies from MHz to GHz to achieve penetration depths from meters to kilometers [156], [157].

The past decade has witnessed the development of drone-based GPR, pioneered by the successful sounding of glacial ice in 2014 using a UAV-based radar system, called G1X [158], which was equipped with a dual-frequency radar operating at approximately 14 and 35 MHz. In 2017, lighter UAV-based radar systems, G1XB and G1XC, with extended endurance of up to 1 hour and 15 minutes (30 minutes longer than earlier systems), were reported to have been deployed in Greenland following earlier missions in Antarctica [159]. During this mission, 35-MHz radar sounder data were collected and compared with previously acquired very-high-frequency (VHF) radar sounder data. The results showed that the 35-MHz sounder not only reproduced the ice-bottom features well detected by the VHF sounder but also identified additional regions of the ice bottom that the VHF sounder failed to capture, due to the effects of surface roughness and temperate ice. Subsequently, in [160], the same systems were employed for sounding temperate glaciers in Russell Glacier. The experiments demonstrated that the high-frequency (14 MHz) and Low-VHF (30–35 MHz) range provided a favorable trade-off between penetration depth and cross-track aperture, effectively suppressing surface clutter. In addition, researchers investigated advanced processing methods to further enhance signal quality. Moreover, a field deployment at Helheim Glacier in the summer of 2022 demonstrated that a Low-VHF radar integrated on a small UAV achieved ice-bed detection capabilities comparable to those of the Twin Otter multi-channel coherent radar depth sounder array onboard manned aircraft [161].

Subsequently, numerous glacier and ice detection missions have been conducted using radar systems mounted on unmanned aircraft [102]. The University of Kansas multi-band instrumentation (14 MHz–38 GHz) has been applied in retrieving key cryosphere parameters such as ice thickness and bedrock topography, and researchers highlighted the significance of field programs and international collaborations, particularly in Chile. In another example, a chirped radar system based on software-defined radio was introduced for deployment on a low-cost and easily transportable fixed-wing UAV. With antennas fully integrated into the UAV's wings, offering a frequency range of 300–450 MHz, preliminary flight tests at an altitude of 120 m demonstrated clear imaging of the lake bed [102].

In recent years, UAV-borne GPR technology for sub-ice detection has advanced significantly, encompassing both hardware and software developments. In terms of hardware, miniaturization and system integration have been the key trends. For instance, [162] introduced a compact and lightweight (0.6 kg) bank of transmit/receive modules for a UAV-borne radar operating at 60–80 MHz, distributed across four antennas and providing approximately 51 dBm output power, 104 dB transmit isolation, and 18 dB receive gain prior to post-amplification. Similarly, [104] detailed the development of an integrated UAV-based GPR system designed for acquiring high-resolution data on glacier internal structures and bed conditions. The authors effectively addressed existing research gaps by combining state-of-the-art components, including a custom GPR controller for long-duration missions, lightweight antennas, a terrain-tracking system, and high-precision differential GPS positioning.

The development of novel, tailored software processing techniques has kept pace. In [104] and [163], standard GPR preprocessing methods—such as background (mean trace) removal, band-pass filtering, and time-varying gain correction—were employed to enhance sub-ice target visibility. Moreover, [115] proposed an advanced data processing workflow tailored for UAV-borne GPR, incorporating a common-reflection-surface-based wavefield separation technique and a weighted-negative-entropy-based velocity estimation algorithm, which was validated using experimental data collected over a frozen Antarctic lake.

Compared with snowpack investigations, glacier and ice-sheet sounding impose substantially different electromagnetic and operational requirements. Existing studies indicate that low-VHF and VHF radar systems (typically 14–150 MHz) provide a more favorable trade-off between penetration depth, clutter suppression, and coherent imaging stability in glacier environments because cold glacier ice exhibits relatively low dielectric attenuation. These systems have demonstrated the penetration scale up to a few kilometers with effective bedrock detection capabilities, particularly in polar ice where liquid-water content remains limited. However, the associated long wavelengths require physically larger antennas, increased transmitted power, and stricter platform stability, thereby imposing significant payload and endurance constraints on UAV platforms. Consequently, fixed-wing radar sounders and low-speed survey strategies are generally more suitable for deep glacier sounding than compact multirotor systems. Furthermore, successful glacier imaging strongly depends on advanced motion compensation, SAR focusing, and clutter suppression techniques to mitigate phase instability induced by UAV motion and rough ice surfaces.

The reviewed studies collectively suggest that glacier-oriented UAV-GPR systems occupy an intermediate regime between shallow snow sounding and highly attenuative permafrost investigations. Frequencies in the tens of MHz range provide improved penetration and reduced volumetric scattering compared with GHz snow radars, but at the expense of reduced resolution and increased antenna size. While these configurations are highly effective in cold, low-loss glacier ice, their performance deteriorates in temperate or water-rich ice due to enhanced attenuation and clutter. Therefore, system configurations developed for glacier sounding cannot be directly generalized to permafrost environments.

**Permafrost Studies**

Mapping the internal and basal structures of permafrost is essential for understanding cryosphere dynamics, hydrological connectivity, and climate-driven degradation [4]. Radar investigations primarily aim to delineate the active layer, permafrost table, intra-permafrost ice bodies, and the sub-permafrost interface separating frozen from unfrozen sediments [164]. However, high-resolution imaging of these structures remains difficult due to strong attenuation, scattering, and dielectric heterogeneity in frozen ground.

Ground-coupled GPR has been extensively applied to map permafrost table depths, ice wedges, and massive ice bodies [165]. These systems provide excellent coupling and centimeter-scale resolution but suffer from limited areal coverage, accessibility constraints, and safety concerns in fragile terrains. Complementary approaches such as electrical resistivity tomography [165], seismic refraction [166], and borehole logging [167] have been used to validate GPR-derived stratigraphy, though they remain logistically demanding in remote polar and alpine regions. Additionally, integrating GPR with optical [168], thermal [169], and LiDAR sensors [170] has enhanced monitoring of thaw and ponding processes in discontinuous permafrost zones [18], highlighting its value within multi-sensor observation frameworks.

UAV-borne GPR offers a rapid, non-invasive alternative for mapping frozen-ground structures where surface access is restricted. As we illustrated in previous subsections, most existing UAV-GPR studies focus on cryosphere analogs, such as snowpack characterization and glacier or ice-sheet mapping, demonstrating the capability of UAV platforms for high-resolution subsurface imaging. Yet, direct detection of subsurface permafrost active layers using UAV-borne GPR remains absent in current literature. This limitation primarily arises from fundamental constraints associated with air-coupled GPR operation. Specifically, air coupling significantly reduces EM energy transmission into the ground, resulting in limited penetration depth and increased signal attenuation [171], [172]. In permafrost environments, increasing UAV flight altitude further exacerbates these issues by enlarging the radar–target propagation distance and expanding the first Fresnel zone. This leads to degraded lateral resolution and increased wavefront curvature mismatch, thereby impairing coherent focusing in migration-based imaging [173]. Such degradation is particularly critical when delineating the permafrost interfaces, where accurate characterization of the active layer boundaries is essential.

Moreover, the increased two-way travel path length ($Z$) introduces stronger geometric spreading losses (proportional to $1/Z^2$ in power), which significantly attenuate backscattered signals from deeper or weakly contrasting subsurface features [174]. This is especially problematic for detecting ice-poor permafrost or thawed zones, where dielectric contrasts are inherently weak, resulting in reduced SNR. Concurrently, the strong dielectric contrast at the air–ground interface produces dominant surface reflections that can obscure weaker subsurface echoes, particularly those associated with the base of the active layer or discrete ice lenses [175]. The extended propagation path also amplifies multipath interference and surface reverberation effects, leading to waveform distortion and reduced separability of closely spaced reflectors. In addition, the enlarged antenna footprint at higher flight altitudes induces spatial averaging over laterally heterogeneous features, such as ice wedges and thaw structures, thereby limiting the detection of fine-scale variability [176]. These challenges are further intensified in rough and vegetated terrains, where surface undulations introduce phase errors and scattering, and vegetation layers contribute additional volumetric scattering and attenuation, increasing clutter and further masking subsurface reflections [26]. Collectively, these factors significantly hinder accurate interface localization, active layer thickness estimation, and reliable characterization of permafrost heterogeneity.

Beyond propagation-related limitations, UAV-borne GPR systems are constrained by payload capacity and onboard power, limiting the deployment of low-frequency antennas and exacerbating the frequency–depth trade-off. Low-frequency operations, required for deep penetration into ice-rich permafrost and for imaging the permafrost base [25], entails large physical dimensions (on the order of half-wavelength dipoles) and higher power capacity to maintain adequate radiation efficiency [177]. These requirements often exceed the payload and energy budgets of typical UAV platforms. Furthermore, due to EM wave attenuation in lossy media [23], higher frequencies provide improved resolution but reduced penetration, particularly in conductive or partially thawed soils with elevated liquid water content [178]. In permafrost settings, this trade-off is especially pronounced: high-frequency systems are suited to shallow investigations, such as the active layer and near-surface ice features, but are ineffective for imaging deeper interfaces or massive ice bodies. Whereas low-frequency systems enable deeper sensing at the expense of resolution and increased system burden. Consequently, UAV-borne GPR configurations must balance penetration depth, resolution, and platform limitations, often resulting in suboptimal performance for comprehensive permafrost characterization.

TABLE 1. MAIN CHARACTERISTICS AND DESIGN TRADE-OFFS OF UAV-BORNE GPR SYSTEMS FOR CRYOSPHERIC APPLICATIONS

| | Application | GPR Model/ Antenna | GPR Tech-nology | Center Freq. /BW (MHz) | Penetration Depth (m) | Flight Altitude (m) | Speed (m/s) | Resolution vs Penetration | Platform Constraint | Recommend-ed Use | Ref., Year |
|---|---|---|---|---|---|---|---|---|---|---|---|
| Snow-pack | Snow depth over sea ice | Exponentially Tapered Slot Antenna | SFCW | 1000 - 6000 | 0.1 - 1 | 5 - 15 | 1 - 3 | High resolution, shallow penetration | Moderate altitude, wide-area coverage | Sea-ice surface and shallow layer mapping | [141], 2021 |
| Snow-pack | Snow thickness | FieldFox/Horn antenna | SFCW | 1500 - 4500 | >0.2 | 5 | N.A. | Very high resolution, shallow penetration | Moderate altitude, limited depth capability | Fine snow layering, near-surface features | [103], 2017 |
| Snow-pack | Snowpack evolution assessment | Radar Systems Inc. Zond | Pulsed | 1500 | >0.7 | 1 | 1.2 | High resolution, shallow penetration | Low altitude, slow speed | Temporal monitoring of snowpack changes | [143], 2022 |
| Snow-pack | Avalanche rescue | Horn antenna | FMCW | 1000 - 4000 | 0.8 | 2 - 4 | 0.4 | High resolution, very shallow penetration | Low speed, precise scanning required | Surface-near victim detection | [154], 2021 |
| Snow-pack | Snow thickness | IDS K2 Unit | Pulsed | 900 | >1 | 7.5 | 5 | Moderate resolution–penetration trade-off | Medium altitude, moderate speed | General snow thickness surveys | [142], 2022 |
| Snow-pack | Snow characterization | Spiral and Vivaldi antennas | FMCW | 700 - 4500 | 1.2 | 8 | 21 | Moderate resolution, limited penetration | High-speed survey; reduced data density | Rapid snowpack characterization | [140], 2020 |
| Snow-pack | Avalanche rescue | RIS One GPR, IDS | Pulsed | 400, 600 | 1.5 | 6 | N.A. | Moderate penetration, moderate resolution | Medium altitude deployment | Shallow-to-mid burial detection | [153], 2008 |
| Snow-pack | Avalanche rescue | Spiral and Vivaldi antennas | FMCW | 950 - 6000 | >1.55 | 1 | 2 - 3 | High resolution, limited penetration | Low-altitude, low-speed operation | Victim localization, shallow burial detection | [138], 2018 |
| Snow-pack | Avalanche rescue | RIS One GPR, IDS | Pulsed | 400 | 4.6 - 5.6 | 6/1 | 4.2 | Deep penetration, coarse resolution | Variable altitude, moderate mobility | Victim detection under thick snow | [152], 2009 |
| Snow-pack | Snow thickness | Radsys | Pulsed | 1000 | >4.8 | 2 - 4 | >2 | High resolution, moderate penetration | Low-altitude UAV, moderate coverage | Snow stratigraphy, active-layer mapping | [137], 2024 |
| Snow-pack | Snow characterization | Radar Systems Inc. | Pulsed | 1000, 75 - 400 | 6 | 2 - 4 | 2 - 3 | Balanced resolution and penetration | Low-altitude UAV, moderate endurance | Snowpack structure and internal layering | [105], 2025 |
| Snow-pack | Avalanche rescue | Radar Systems Zond-12e Drone 500A Lite | Pulsed | 500 | 6 - 8 | 1 - 2 | 0.5 | Moderate penetration, reduced resolution | Low speed, low altitude, high precision | Deep burial detection in rescue missions | [144], 2023 |
| Snow-pack | Snow depth | Spiral and Vivaldi antennas | FMCW | 950 - 6000 | >7 | 1 | N.A. | Moderate resolution–depth balance | Low altitude, stable acquisition | Snow depth estimation in heterogeneous layers | [139], 2019 |
| Snow-pack | Snow detection, Glacier crevasses, Avalanche rescue | EKKO 1000 | Pulsed | 900 | 10 | >5 | 4.2 - 8.3 | Moderate resolution, deep penetration | Higher altitude, wide-area survey | Crevasse detection, deep snow analysis | [148], 2003 |
| Ice forms | Subglacial velocity estimation | ZOROY | Pulsed | 400 | >4 | 5 | 2.5 | Moderate resolution–penetration balance | Low-altitude UAV, stable flight required | Ice interface tracking | [115], 2024 |
| Ice forms | Ice depth | Zond Aero LF | Pulsed | 50, 100, 200 | >10 | 5 | N.A. | Deep penetration, low resolution | Low altitude UAV, efficient coverage | Ice depth estimation | [163], 2024 |
| Ice forms | Glaciers mapping | Utsi Electronics Ltd. (UK) | Pulsed | 80 | 100 | 5 | 4 | Deep penetration, coarse resolution | Low altitude, moderate coverage | Glacier thickness mapping | [104], 2023 |
| Ice forms | Glaciers imaging | Resistively loaded dipole antenna/ tapered planar dipole antenna | Pulsed | 14/35 | >300, 500, 800 | 122 - 183 | 30 | Very deep penetration, very low resolution | Heavy system, high-altitude airborne survey | Ice thickness, subglacial structure | [158], 2014; [159], 2017; [160], 2018 |
| Ice forms | Ice-bed detection | Egyptian Axe Dipole (EAD) | Pulsed | 37.3 | >700 | 100 | N.A. | Extreme penetration, coarse resolution | High-altitude, large platform required | Ice-bed interface mapping | [161], 2025 |
| | Ice sheets | Bowtie-style antenna | Pulsed | 300 - 450 | N.A | N.A | N.A. | Moderate penetration, moderate resolution | General-purpose configuration | Ice sheet internal structure | [102], 2022 |
| | Ice sheets | Dipole antenna | Pulsed | 60 - 80 | N.A | 610 - 2438 | N.A. | Deep penetration, low resolution | Very high altitude, large platform | Large-scale ice sheet profiling | [162], 2025 |
| Permafrost | Permafrost layers detection | Zond Aero LF dipole antenna | Pulsed | 100, 150, 300 | 20 | 1.15 | 0.75 | Moderate penetration, reduced resolution | Extra-lightweight, Low-altitude UAV, stable flight required | Permafrost in alpine terrain | [183], 2025 |

To address these limitations, several implementation-oriented strategies can be considered. First, multi-sensor integration offers a viable pathway to compensate for the intrinsic limitations of UAV-GPR. Complementary geophysical techniques, such as refraction seismic, electrical resistivity tomography (ERT), and EM induction, can provide additional constraints on subsurface structure, ice content, and bedrock properties, thereby enhancing interpretability [179]. Furthermore, the integration of UAV-borne GPR with LiDAR [170] and thermal imaging [169] can improve the characterization of surface morphology, moisture distribution, and ice heterogeneity. High-resolution topographic mapping and vegetation indices (e.g., greenness maps) can also be leveraged to identify optimal survey areas and provide critical contextual information, particularly in complex terrain and densely vegetated environments [168], [180].

In terms of hardware constraints, advances in lightweight and broadband antenna design present promising opportunities to mitigate payload and power limitations [104]. Emerging antenna technologies aim to reduce size and weight while maintaining sufficient bandwidth and radiation efficiency, thereby improving compatibility with UAV platforms [181]. In addition, recent developments in compact, modular GPR systems constructed from off-the-shelf components demonstrate the feasibility of lightweight, energy-efficient designs suitable for autonomous and long-term deployment [182]. Collectively, these strategies provide practical avenues to enhance the detectability and reliability of UAV-borne GPR for permafrost investigations.

Sub-permafrost detection using UAV-borne GPR remains largely unexplored, however, advances in low-frequency airborne antennas, flight control precision, and multi-sensor integration suggest its strong future potential for large-area, high-resolution mapping of frozen-ground stratigraphy. A pioneering example is the UAV-GPR survey conducted by the Surface Processes and Remote Sensing Laboratory (SPaRS-Lab), University of New Brunswick, in the vegetation-covered alpine terrain of the Chic-Choc Mountains (Gaspé Peninsula, Canada), where conventional surveys are impractical. The mission integrated photogrammetry and LiDAR with a low-frequency (100–300 MHz) UAV-mounted Zond Aero LF GPR (SPH Engineering, ~0.9–1.2 kg), achieving penetration depths of up to ~20 m at 100 MHz. Operating at ~1.15 m AGL and 0.75 m/s, the UAV acquired over 1.5 km of GPR data within one hour. The dataset is expected to be publicly released, with related peer-reviewed publications forthcoming [183]. In contrast, ground-based GPR investigations are typically constrained to localized measurements [184], [185]. For instance, a study near Barrow, Alaska, USA, conducted a 27.5 m survey using a 500 MHz antenna, where the relatively high frequency likely reflects payload limitations, labor-intensive field deployment, and logistical challenges such as steep terrain, long access routes, and dense vegetation, and detected ice wedges at depths of approximately 4.53 m [185]. This quantitative comparison underscores the advantages of UAV-borne GPR, particularly its ability to carry targeted designed, lightweight lower-frequency antennas, and to achieve extended survey coverage even in challenging environments.

The reviewed literature indicates that permafrost investigations require a fundamentally different system-design philosophy from snowpack or glacier surveys. Successful deployment depends not only on lower operating frequencies and motion-compensated processing, but also on strict low-altitude flight control, multi-sensor integration, and careful trade-offs among penetration depth, resolution, payload capacity, and energy consumption. In contrast to glacier radar sounding, where low dielectric loss enables deep penetration, permafrost environments are often dominated by strong attenuation and clutter, making detection performance highly sensitive to soil moisture, vegetation cover, and surface roughness. Therefore, deployment-oriented strategies—including lightweight broadband antennas, LiDAR-assisted terrain following, and integrated geophysical sensing—become essential for reliable UAV-borne permafrost characterization.

Overall, the reviewed studies demonstrate that the optimal UAV-borne GPR configuration in cryosphere environments is strongly condition dependent and governed by the coupled effects of dielectric properties, target depth, platform constraints, and environmental variability. Snowpack investigations generally favor GHz-band ultra-wideband systems for high-resolution shallow imaging, whereas glacier radar sounding relies on low-VHF radar to achieve much deeper penetration in low-loss ice. In contrast, permafrost investigations require lower frequencies, stricter flight stability, and multi-sensor integration to mitigate strong attenuation and heterogeneous scattering. These differences indicate that no universal UAV-GPR architecture is suitable for all cryosphere applications. Instead, effective deployment requires application-specific co-design of antenna geometry, center frequency, bandwidth, flight altitude, and processing workflow according to the electromagnetic characteristics and operational constraints of the target environment, which will be detailed in the following sections.

**Operational Strategies for UAV-Borne GPR in Cryosphere Environments**

Flight altitude and speed are tightly interrelated parameters in UAV-borne GPR surveys, as they jointly affect radar signal coupling, SNR, imaging resolution, and survey productivity. Optimal settings depend on terrain morphology, system response delay, and the EM properties of the near-surface layer, such as air-ice and air-snow interfaces.

A critical feature in such surveys is the terrain-tracking mode, such as the True Terrain Following (TTF) navigation system developed by SPH Engineering (Latvia) [104], which adjusts the UAV's altitude to maintain an approximately constant height above the glacier surface. However, practical limitations exist: typical 1s delays between radar altimeter measurements and altitude correction can produce significant horizontal offsets over abrupt terrain features. For instance, during a 2022 survey at Otemma Glacier, Switzerland, a UAV flying at 4 m/s with a programmed altitude of 5 m experienced a 5 m horizontal offset when passing a lateral moraine, a situation that could have resulted in a crash if the UAV had flown lower or faster [104].

Lower altitudes, where UAV flies lower than 1-2 m AGL, improve antenna coupling with the sub surface by minimizing the air gap, thus enhancing energy transmission into the ice or snow [104]. This typically yields stronger reflections and more continuous subsurface interfaces. However, maintaining extremely low altitudes is risky in rugged or crevassed terrains, as small topographic variations can exceed the UAV's terrain-tracking correction limits. Such conditions require reduced flight speeds to ensure stable TTF response and prevent delayed corrections that might lead to surface collisions. Consequently, very low-altitude, slow-speed configurations are generally reserved for detailed, small-scale profiling or internal-layer mapping over relatively smooth surfaces [104], [105].

At intermediate altitudes (2–5 m AGL), UAVs can safely achieve higher speeds (2–4 m/s) with minimal loss of signal fidelity. Experiments demonstrated that SNR remains high within this range, while internal reflections and bed returns are still clearly resolved [175]. This altitude-speed combination represents an effective trade-off between safety, data quality, and coverage rate—especially valuable for glacier-scale mapping or time-lapse (4D) surveys requiring repeated flight lines. In addition, moderate altitudes minimize the influence of UAV pitch and roll, maintaining a consistent incidence angle and improving cross-line coherence [105], [115].

Higher altitudes (>5 m AGL) permit faster coverage (≥5 m/s) and are advantageous for initial reconnaissance or wide-area mapping. However, the radar signal undergoes increased attenuation and geometric spreading in air, reducing the amplitude of subsurface reflections. The air layer also introduces velocity discontinuities, leading to slight distortions of diffraction hyperbolas [175] and time delays that complicate migration or velocity analysis. To correct for these effects, a layered air–ice velocity model or ray-based correction must be applied during post-processing [78], [115],[175].

Speed also affects data sampling density and system stability. Slower speeds improve spatial sampling, capturing fine-scale subsurface structures and preserving diffraction integrity [105], [115]; however, they substantially decrease the area covered per flight and increase battery depletion speed. Higher speeds, by contrast, enhance operational efficiency but may introduce minor positional inaccuracies and increase motion-induced noise. For glaciological applications focusing on layer mapping rather than fine-scale scattering analysis, moderate to high speeds (2–4 m/s) are generally acceptable, as the dominant subsurface interfaces remain well resolved [159].

In complex terrain, flight planning must incorporate slope direction and prevailing wind conditions. Downslope or slope-parallel trajectories mitigate altitude drift caused by UAV pitch adjustments and improve stability in radargram continuity [105]. Moreover, adaptive flight speed control—reducing velocity during steep ascents or near obstacles—can further optimize data quality without compromising safety.

Weather conditions play a crucial role in determining the safety, stability, and data quality of UAV-borne GPR operations. Ideal flight conditions are clear and calm days, free from precipitation, fog, and strong winds [181]; however, such circumstances are rarely encountered in polar or mountainous environments. In practice, flight planning typically relies on high-resolution meteorological forecasts, such as those provided by AEMET [182], which offer predictions of cloud cover, humidity, wind speed and gusts, precipitation, and temperature. Wind strongly affects UAV stability, flight speed, and GPS positioning accuracy, leading to potential errors in the spatial registration of A-scan points and, consequently, in the resolution and accuracy of subsurface imaging [115]. Precipitation and fog further degrade radar performance by attenuating EM signals and interfering with sensor calibration. To minimize these effects, field tests are generally scheduled under clear weather and low-wind conditions, with real-time monitoring of atmospheric parameters using onboard or nearby meteorological sensors [115]. Low temperatures also reduce battery efficiency and limit flight endurance, making proper thermal insulation and optimized energy management essential for sustained operations in cold environments [5]. Therefore, careful pre-flight assessment of meteorological data and adaptive flight planning are critical to ensure both UAV safety and the reliability of GPR data acquisition in harsh environmental settings [181].

To provide quantitative mission-design guidance, the key acquisition parameters of UAV-borne GPR systems—namely flight altitude $Z_0$, platform tangential velocity $v_r$, PRF, antenna 3 dB beamwidth $\theta_{3dB}$, wavelength $\lambda$, and the number of A-scan samples $num$ within the synthetic aperture length $L_s$—can be constrained through the following relationships:

$$PRF \geq \frac{4v_r}{\lambda} sin\left(\frac{\theta_{3dB}}{2}\right), \quad (17)$$

$$num \geq \frac{8Z_0 tan\left(\frac{\theta_{3dB}}{2}\right)}{\lambda} sin\left(\frac{\theta_{3dB}}{2}\right). \quad (18)$$

These expressions establish the minimum sampling requirement along-track (via PRF) and within the synthetic aperture (via $num$) to satisfy spatial sampling criteria and suppress aliasing in SAR-based GPR imaging. Notably, they explicitly link platform motion parameters ($v_r, Z_0$) and antenna characteristics ($\theta_{3dB}, \lambda$) to data acquisition settings, thereby enabling principled survey design. The detailed derivation of these relationships is provided in Appendix.

In addition, the temporal sampling of the received signal is governed by the relationship between the system bandwidth (or effective sampling rate) $BW$, the number of samples $N$, and the total time window $T$:

$$T = \frac{N}{BW}. \quad (19)$$

This relationship defines the maximum observable two-way travel time and, consequently, the achievable detection depth for a given EM wave velocity $v_\epsilon$ in the medium, which is primarily determined by its dielectric constant $\varepsilon$, $v_\varepsilon = c/\sqrt{\varepsilon}$, where $c$ is EM wave velocity in air. It further establishes a direct trade-off between the temporal (range) resolution (governed by the system bandwidth $BW$ and medium dielectric constant $\varepsilon$) and the observation window length (controlled by the number of samples $N$).

The lateral (cross-range) resolution can be approximated as [45], [183]:

$$\Delta_x \approx 0.886 \frac{\lambda}{2L_s}\left(Z_0 + \frac{z}{\sqrt{\varepsilon}}\right), \quad (20)$$

where $z$ denotes the target depth. These parameters must be jointly configured according to the expected target depth range, characteristic spatial scale and dielectric constant.

The received power $P_r$ can be expressed as [184], [185]:

$$P_r = P_t \left(\frac{\lambda_{air}}{4\pi}\right)^2 \frac{RG_a^2 T^2 L_A G_{sys}}{4\left(Z_0 + \frac{z}{\sqrt{\varepsilon}}\right)^2}, \quad (21)$$

where $P_t$ is the transmitted power, $\lambda_{air}$ is the radar wavelength in air, $R$ is the reflection coefficient at the medium/target interface, which is related to the dielectric constant of the subsurface medium, $G_a$ is the antenna gain, T is the transmission loss at the air/medium interface $T = 1/(1 - R^2)$. $L_A$ is two-way propagation loss through the medium $L_A = \exp(2\alpha z)$, where $\alpha$ is the dielectric loss in nepers per meter. $G_{sys}$ represents system-level gains, including hardware efficiency and processing gain. To ensure reliable detection, the received signal power and the system noise floor $P_n$ satisfy a minimum system SNR requirement:

$$\frac{P_r}{P_n} \geq SNR_{sys_min}. \quad (22)$$

This condition explicitly links antenna gain, transmitted power, propagation losses, and target dielectric properties to the maximum detectable depth, thereby providing a quantitative basis for mission design and performance prediction.

To demonstrate the practical application of (17)–(22), an illustrative mission-design example for detecting shallow ice structures in Antarctic environments is presented. Consider a UAV-borne GPR survey designed to image a near-surface ice crack buried at approximately $Z = 1$ m depth within compacted snow/firn layers, where the effective dielectric constant is assumed as $\varepsilon$=3.15. A low-altitude survey configuration with flight altitude $Z_0$ =5 m and platform velocity $v_r$ =2 m/s is adopted to balance footprint coverage and coherent imaging stability under windy polar conditions. Assuming a center frequency of 400 MHz ($\lambda$≈0.75 m in air) and an antenna 3 dB beamwidth of $\theta_{3dB} = 60^\circ$, (17) yields a minimum $PRF$ requirement of approximately 5.3 Hz to avoid Doppler aliasing during SAR processing. Considering motion uncertainties and coherent integration requirements, a practical $PRF$ of 50–100 Hz is recommended. From (18), the minimum number of A-scan samples within the synthetic aperture is approximately 31, ensuring sufficient along-track spatial sampling for SAR image focusing. For depth coverage and range resolution analysis, the EM propagation velocity in ice is approximately $v_\epsilon = \frac{c}{\sqrt{3.15}} \approx 1.69\times10^8$ m/s. Assuming a bandwidth of 300 MHz, the achievable vertical resolution is on the order of several tens of centimeters, which is sufficient for resolving meter-scale ice structures. Using (20), the corresponding cross-range resolution is estimated to be approximately 0.3–0.5 m, depending on the effective synthetic aperture length $L_s$, detailed in Appendix $(A-11)$. Furthermore, according to the radar link-budget relationships in (21)–(22), and transmitted power is set as $P_t =$ 1W, antenna gain $G_a = 6$ dBi, and a representative reflection coefficient of $R = 0.2$ to model moderate dielectric contrasts between snow/firn layers and shallow ice structures, the transmission loss $T$ is calculated around 1.04. For shallow ice targets, propagation attenuation is assumed small ($L_A \approx 1$), yielding an estimated received signal power of approximately $P_r \approx 1\times10^{-5}$ W (~-20 dBm). Therefore, the reliable target detection can be achieved under the proposed survey configuration, provided that the received signal power satisfies the minimum system SNR requirement relative to the system noise floor $P_n$. This example demonstrates how the proposed framework can quantitatively guide the joint selection of UAV flight parameters, radar configurations, and acquisition settings according to target depth, dielectric properties, and desired imaging performance in cryosphere environments.

## UAV-Borne GPR Experiment for Characterizing Frozen Ice at Mochou Lake, Antarctica

This section presents the results obtained by processing data collected with the UAV-borne GPR (Fig. 4) reported by Luo et al [115]. The experimental system integrates an IGPR-30 GPR, manufactured by Dalian Zoroy company, with a DJI M600 Pro hexacopter platform. The radar unit operates at a center frequency of 400 MHz and employs a compact bowtie antenna featuring a fixed, non-adjustable beamwidth[115]. The DJI M600 Pro, a medium–large multirotor UAV with a maximum takeoff weight of approximately 6 kg, provides the payload capacity and flight stability necessary for reliable radar operation [115].

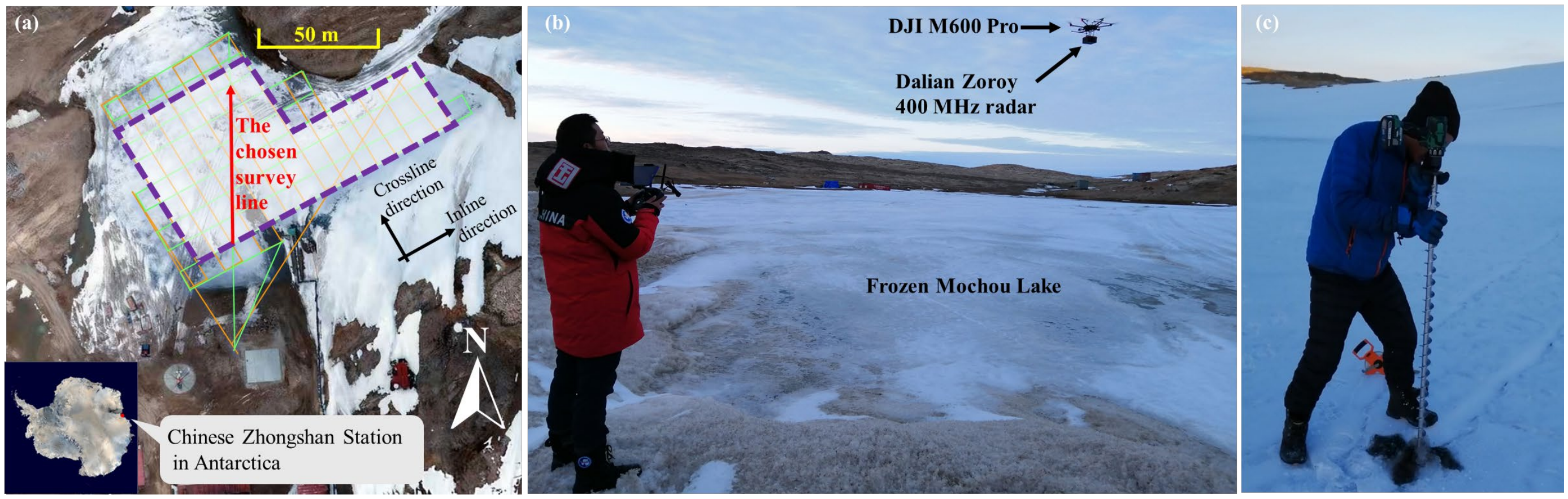

**Fig. 4.** (a) Survey layout: green/orange solid grids denote the surveyed area; the red arrow indicates the example survey line for processing illustration; the purple dashed frame marks the 3D reconstruction region. (b) UAV operation during data acquisition. (c) Drilling procedure for ground-truth ice-thickness measurement.

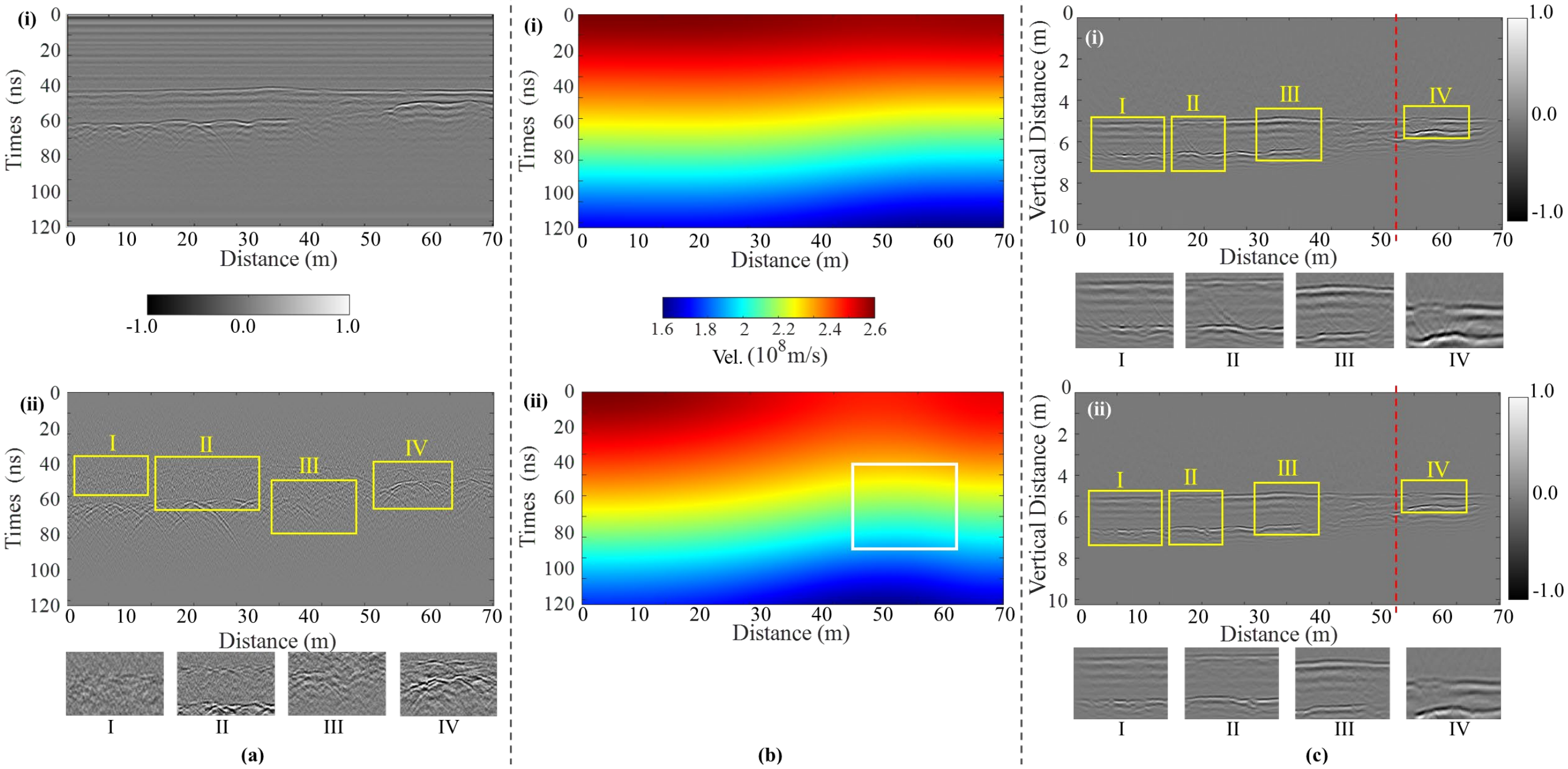

**Fig. 5.** UAV-GPR data through data processing workflow. (a) Normalized UAV-borne GPR data. (a-i) Raw data. (a-ii) Data after wavefield separation. (b) EM wave velocity estimation results. (b-i) conventional diffraction focusing-based method. (b-ii) WNE-based method. (c) Imaging results using f-k migration based on velocity results of (c-i) conventional diffraction focusing-based method and (c-ii) WNE-based method.

The data was collected on the frozen Mochou Lake, located next to the Chinese Zhongshan Station in East Antarctica, as shown in Fig. 4. The UAV-borne GPR operated at an altitude of around 5 m above the ice surface (Fig. 4(b)) to ensure sufficient signal return and avoid aerodynamic disturbance [186]. The UAV followed the solid grids pattern illustrated in Fig. 4(a), maintaining a flight speed of about 2.5 m/s and a data acquisition rate of 100 traces/s. Each A-scan comprised four averaged traces, resulting in a spatial sampling interval of 0.1 m, which is below half the wavelength at 400 MHz (0.375 m) and thus prevents Doppler aliasing while ensuring high-resolution imaging [91]. The time-sampling interval was 0.12 ns, with 1024 samples recorded per trace [115]. A 70 m survey line, indicated by the red arrow in Fig. 4(a), was selected from the scanning grid for detailed data processing and analysis. As shown in the original UAV-borne GPR B-scan image (Fig. 5(a)), strong reflected signals dominate the waveform, completely obscuring the diffraction patterns. A pronounced wavefield aliasing effect is particularly evident within the 40–50 ns and 60–70 ns time windows. These characteristics of the raw data significantly reduce the detectability of radar diffraction signals, leading to large errors in conventional velocity analysis and severely limiting the accuracy of internal ice-structure interpretation.

The raw radargram shown in Fig. 5(a-i) was processed following the workflow described in Section III. The raw data were first processed through a series of preprocessing steps, after which the resulting dataset was subjected to CRS–based wavefield separation method [115]. As illustrated in Fig. 5(a-ii), this approach effectively suppresses strong reflection components, isolating clear diffraction signals. For detailed visualization, four representative zones were enlarged. In Zone I, the air–ice interface reflection is efficiently removed while weak-amplitude diffractions are well preserved. In Zones II–IV, high-resolution diffraction features were successfully extracted, and the separation of strong reflection components did not compromise the integrity of the diffraction signals, providing reliable input for subsequent diffraction-focusing–based velocity analysis.

Velocity estimation was performed using the WNE-based method [115]. Fig. 5(b) presents a comparison of the EM wave velocity fields predicted before and after applying the weighting scheme. Although the two results exhibit similar overall trends, local discrepancies are evident, which can influence the subsequent subsurface imaging quality. For instance, within the region highlighted by the white box, the velocity boundary obtained using the weighted method shows noticeably higher curvature compared to the unweighted result. Between 50 m and 60 m, a distinct convex feature appears, corresponding closely to a hyperbolic diffraction observed in the Zone IV in Fig. 5(a-ii). This correspondence indicates that the weighting function effectively calibrates diffraction features, thereby improving the local velocity estimation. Such differences are expected to directly affect the imaging accuracy of the corresponding subsurface zones.

Fig. 5(c) presents a comparison between the imaging results obtained using the proposed WNE velocity analysis method and the conventional unweighted approach. Both methods successfully reconstructed the upper and lower interfaces of the ice body; however, significant differences are evident in imaging quality. Compared with the unweighted result in Fig. 5(c-i), the WNE-based approach (Fig. 5(c-ii)) produces a smoother and more continuous interface. In Zones I–IV, which include both flat and inclined geometries, the unweighted method exhibits defocused hyperbolic tails and noticeable over-migration artifacts, whereas these anomalies are effectively suppressed in the WNE-based result. The continuity of reflection phase axes is substantially improved. In Region IV (corresponding to the white frame in Fig. 5(b)), the convex feature appears smoother, and over-migration effects are markedly reduced. These comparisons demonstrate that the weighted negative-entropy velocity analysis enhances velocity-picking accuracy, thereby significantly improving the quality and reliability of two-dimensional subsurface reconstructions.

From the perspective of SNR improvement power, which is a key metric for assessing the efficiency of subsurface focusing methods [115], the proposed WNE-based approach exhibits superior performance. As shown in Fig. 5(a), a distinct hyperbolic diffraction centered at approximately 60 ns is observed along the A-scan indicated by the red-dotted line in Fig. 5(c). For this target, the return power obtained using the WNE-based method reaches 21.13 dB above the noise floor, notably exceeding the 19.69 dB improvement achieved by the conventional f–k migration without weighting [27]. The result demonstrates that the proposed method more effectively concentrates diffracted energy at its true spatial origin, thereby achieving a significant SNR enhancement, as shown in Fig. 6. Similar results were consistently observed across other survey lines, as reported in [115].

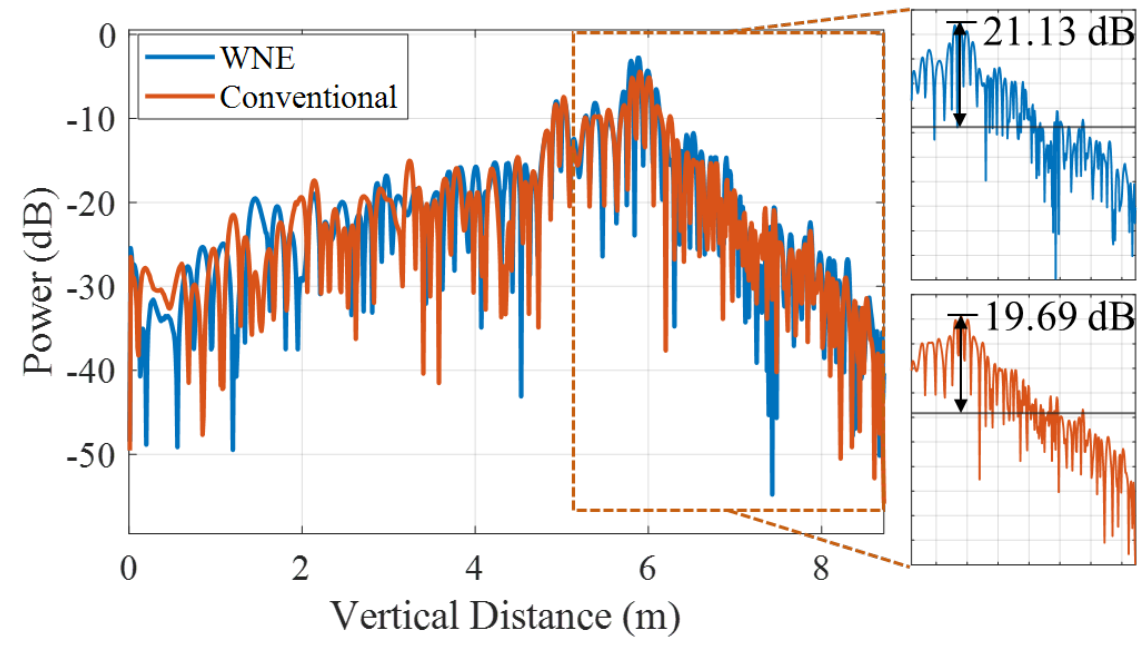

**Fig. 6.** SNR improvement power comparison between red-dotted A-scans in Fig. 5(c).

Fig. 7–Fig. 9 provide complementary 3D views of the processed frozen ice layer dataset collected over Mochou Lake. Fig. 7 shows a 3D view of selected intersecting inline and crossline profiles. In both profiles, the upper and lower ice surfaces are clearly delineated, and their geometries are consistent across the intersections, which occur at approximately 5.04 m and 6.33 m under the UAV flight line. This visualization serves as a crossover-based validation, confirming the consistency of the 2D profiles produced by the UAV-borne GPR acquisition and processing workflow. Fig. 8 provides a detailed 3D view of the frozen ice layer, showing that it spans more than 80 and 60 m along the inline and crossline survey directions, respectively. Because the upper and right-hand boundaries of the survey grid in Fig. 4(a) lie close to the shoreline, the corresponding survey lines exhibit merged reflections where the lower ice interface becomes indistinguishable from the upper surface. These subsets of data were removed during preprocessing. The area used for 3D reconstruction, corresponding to Fig. 8, is indicated by the purple dashed frame in Fig. 4(a).

The distances from the UAV flight line (considered at a constant height as in Fig. 2) to the upper and lower ice surfaces are shown in Fig. 9(a) and (b), respectively. The upper surface lies within the range of 4.78–5.42 m, while the lower surface ranges from 5.40–6.72 m, as derived from the 3D reconstruction. These values are consistent with the expected undulation of the ice surface. The ice thickness distribution of the covered area is illustrated as Fig. 9 (c), with an average depth equals to 1.09 (range of approximately 0.2–1.6 m). A drilling operation was performed at the purple validation point indicated in Fig. 8 and Fig. 9(a)(b)(c) to obtain ground-truth of ice thickness. The drilling process is shown in Fig. 4(c), and the measured thickness was approximately 0.68 m, as reported in Fig. 9(d). The normalized travel-time signal extracted from Fig. 8 at this location, shown as the red line in Fig. 9(e), yields an estimated ice thickness of 0.65 m, with the upper and lower interfaces located at 5.20 m (purple circle) and 5.85 m (yellow circle) beneath the UAV flight line, respectively. This corresponds to a depth-estimation error of only 0.03 m for the WNE-based method, demonstrating the reliability of the UAV-borne GPR results. In contrast, the conventional unweighted strategy, shown as the blue dotted line in Fig. 9(e), identifies the upper and lower ice interfaces at 5.20 m (purple circle) and 5.83 m (yellow circle), respectively, resulting in an estimated ice thickness of 0.63 m and a depth-estimation error of 0.05 m. The underestimated thickness likely results from defocused hyperbolic tails and noticeable over-migration artifacts after migration, as shown in Fig. 5(c)(i), which reduce the vertical separation in the A-scan plot. This comparison demonstrates the improved accuracy of the WNE-based approach.

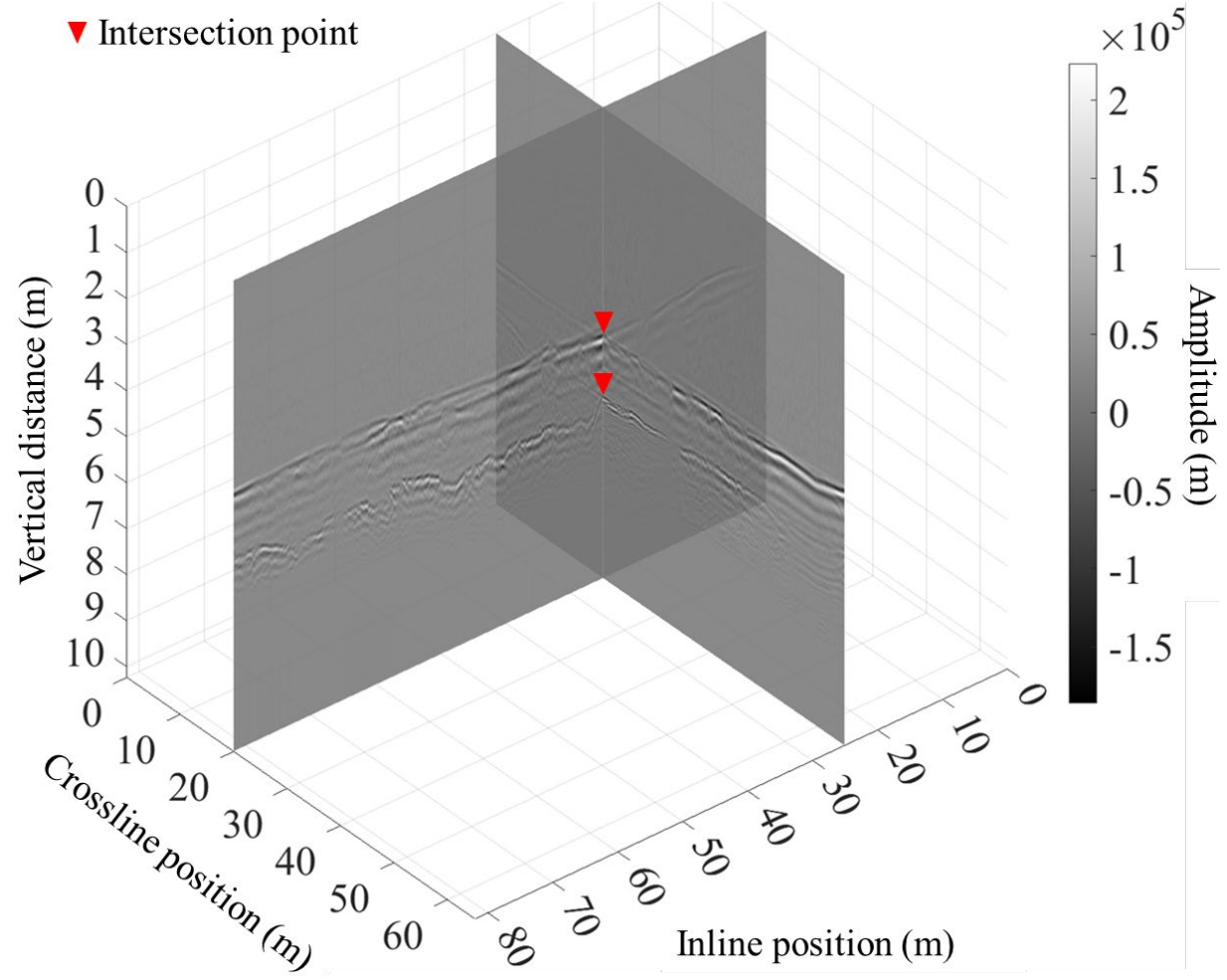

**Fig. 7.** 3D view of a selected intersecting inline and crossline profile.

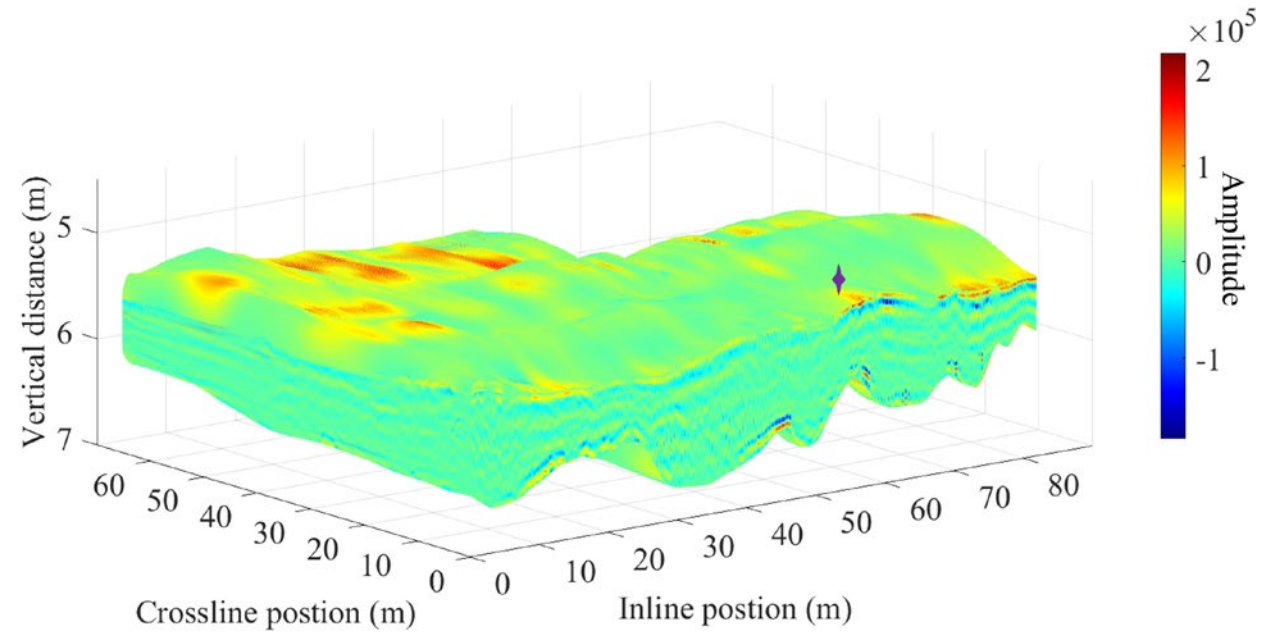

**Fig. 8.** 3D view of the frozen ice layer.

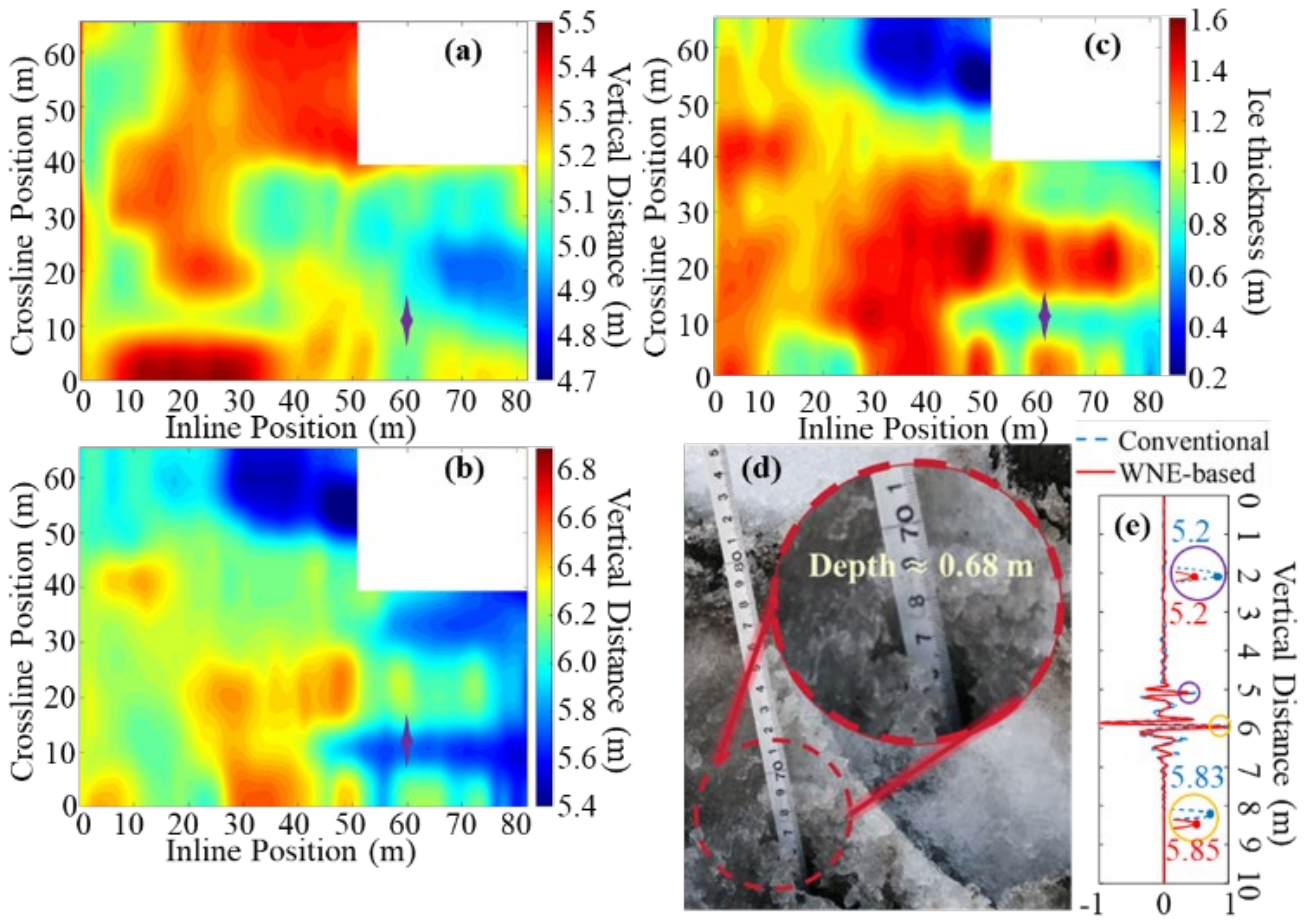

**Fig. 9.** The vertical distances from the UAV flight line to the (a) upper and (b) lower ice surfaces. (c) The ice thickness distribution of the covered area. (d) Measured thickness of approximately 0.68 m at the validation point. (e) Normalized travel-time signal extracted at the same location using the conventional method [27] and our WNE-based strategy [115].

## Challenges and Future Trends

### Quantitative Sensitivity of Phase Stability to Environmental Variability

Despite its proven capability for rapid, non-contact shallow subsurface mapping in extreme environments, UAV-borne GPR face a range of technical, operational, and methodological challenges. Compared with ground-coupled systems, UAV-mounted configurations inherently suffer from increased energy loss and imperfect ground coupling at the air–snow or air–ice interface. These coupling losses reduce the effective SNR and penetration depth, particularly in heterogeneous snow or firn layers where dielectric contrast is high [105]. The elevated antenna geometry also enlarges the Fresnel zone and broadens the radar footprint, degrading lateral resolution and blurring small-scale features such as ice lenses or internal layering. Although low-altitude flight paths can mitigate these effects, maintaining stable, near-surface operation requires precise flight control and robust terrain-following mechanisms, especially under strong winds and variable topography [105].

Platform-induced perturbations, arising from strong winds and variable topography, together with medium-dependent EM variability, further degrade UAV-borne GPR performance, including imaging quality, depth estimation, and subsurface property inversion. Variations in snow/ice density and temperature gradient modify the relative permittivity $\varepsilon$, thereby altering wave velocity and propagation paths, while wind-induced UAV motion perturbs the radar position (along-track-position $A_r$, altitude $Z_0$) and observation geometry. These effects can be consistently interpreted as path errors. First, propagation path specifically adopted in the cryosphere setting is expressed as

$$\hat{Z}_{td} = Z_0\sqrt{1+\left(\frac{A_p - A_r}{Z_0 - \frac{z}{\sqrt{\varepsilon}}}\right)^2} - \sqrt{\varepsilon}z\sqrt{1+\left(\frac{A_p - A_r}{\sqrt{\varepsilon}\left(Z_0 - \frac{z}{\sqrt{\varepsilon}}\right)}\right)^2} \quad (23)$$

where $(A_p, z)$ is the subsurface targets position [183], [187]. Then, perturbations in $\varepsilon$, $A_r$, and $Z_0$ yield a modified path $\hat{Z}_{td}{}'$:

$$\hat{Z}_{td}{}' = (Z_0 + \Delta Z_0)\sqrt{1+\left(\frac{A_p - (A_r + \Delta A_r)}{(Z_0 + \Delta Z_0) - \frac{z}{\sqrt{\varepsilon + \Delta\varepsilon}}}\right)^2}$$
$$-(\sqrt{\varepsilon + \Delta\varepsilon})z\sqrt{1+\left(\frac{A_p - (A_r + \Delta A_r)}{(\sqrt{\varepsilon + \Delta\varepsilon})\left((Z_0 + \Delta Z_0) - \frac{z}{\sqrt{\varepsilon + \Delta\varepsilon}}\right)}\right)^2} \quad (24)$$

with the corresponding path error $\Delta\hat{Z}_{td} = \hat{Z}_{td} - \hat{Z}_{td}{}'$, leading to a phase error

$$\Delta\phi = \frac{4\pi f_c \Delta\hat{Z}_{td}}{c} \quad (25)$$

where $f_c$ is the center frequency, $c$ is the EM wave velocity in air. Since performance is directly governed by phase consistency($\Delta\phi < \pi/4$), these coupled uncertainties can result in defocusing and geometric distortion if not properly compensated [45], [188]. Finally, ensuring stable system performance therefore requires constraining variations in $\varepsilon$, $A_r$, and $Z_0$ within tolerances defined by (23)–(25).

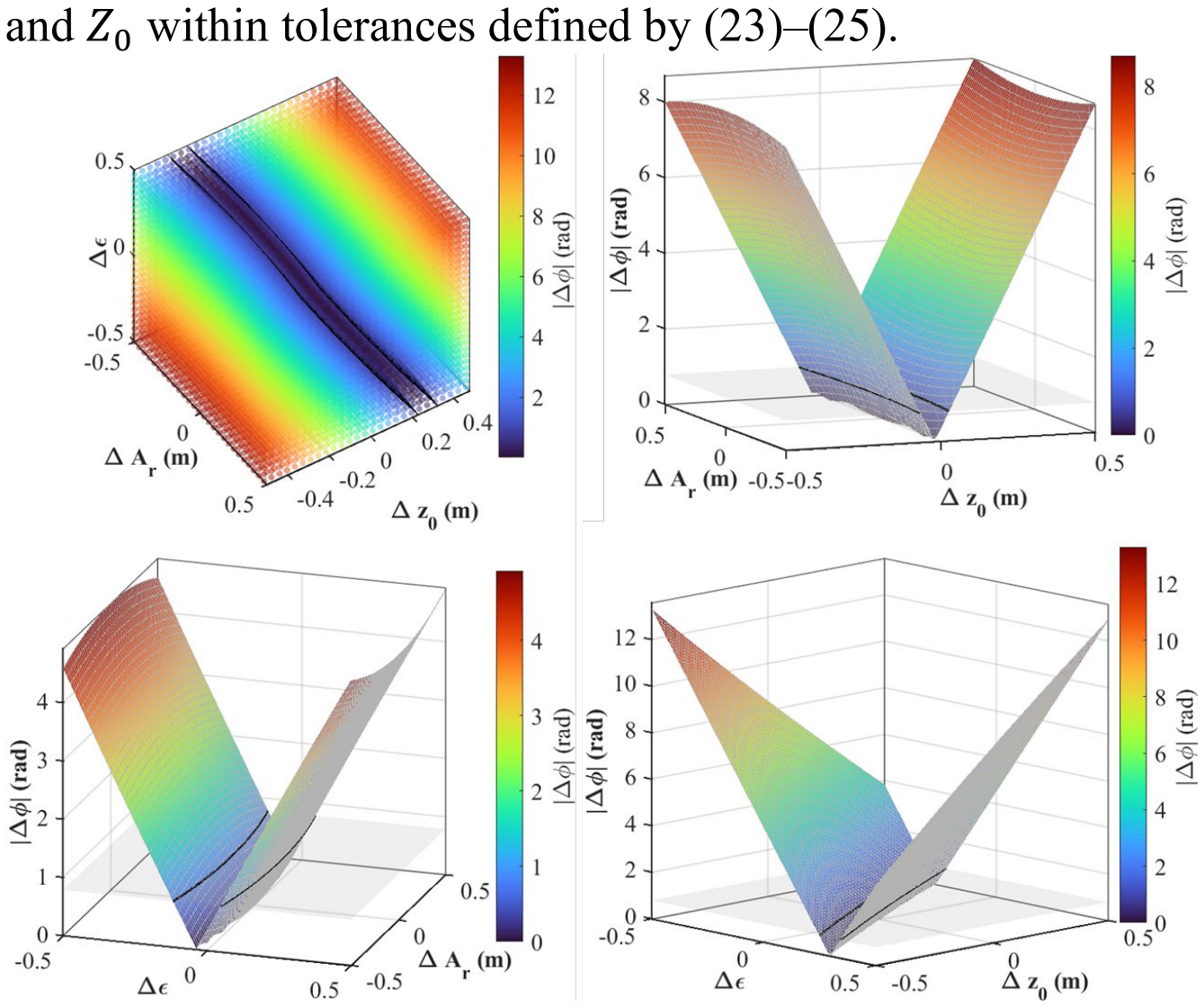

**Fig. 10.** $\Delta\phi$ under variations of key parameters: (a) joint dependence on $A_r$, $Z_0$, and $\varepsilon$, where the region satisfying (25) is bounded by the black surfaces; (b) $\varepsilon$ =3.15; (c) $Z_0$=5 m; (d) $A_r$ =0. Region below gray slice (values below the black curve) denotes conditions under which $\Delta\phi$ remains below the threshold, indicating negligible impact on the final UAV-GPR imaging performance.

According to (23)–(25), a quantitative sensitivity analysis of UAV-borne GPR phase stability under environmental variability is conducted using the experimental scenario described in Section V. A buried target located at (0 m, -2 m) is considered, while the radar platform is positioned at (0 m, 5 m) when directly above the target. The dielectric constant of ice is assumed to be $\varepsilon$=3.15, a commonly adopted value for Antarctic glacial ice [189], [190]. To model environmental perturbations in cryosphere environments, wind-induced variations in both along-track position $A_r$ and flight altitude $Z_0$ are assumed within the range of [−0.5 m, 0.5 m], corresponding to 500% of the nominal along-track A-scan spacing (0.1 m) and 10% of the nominal flight altitude (5 m), respectively. In addition, dielectric constant $\varepsilon$ fluctuations within ±0.5 are considered to account for temperature gradients and snow/firn density heterogeneity commonly encountered in Antarctic ice [189], [190].

The resulting phase-error distributions are illustrated in Fig. 10(a)–(d). As shown in Fig. 10(a), the phase error $\Delta\phi$ is jointly governed by the coupled variations of $A_r$, $Z_0$, and $\varepsilon$. The two black boundary surfaces denote the tolerance region satisfying the phase-stability criterion in (25), indicating that only when the variations of these three parameters remain within this region can the phase error be kept below the threshold, thereby ensuring negligible impact on the final imaging results. To further evaluate parameter sensitivity, additional analyses are performed by constraining one parameter while allowing the remaining two to vary. The corresponding results are presented in Fig. 10(b)–(d), where $\varepsilon = 3.15$, $Z_0$=5 m and $A_r = 0$ are separately fixed. In these cases, the phase error remains below the tolerance boundary indicated by the black curves, demonstrating that accurate control or prior estimation of any one parameter can substantially relax the overall sensitivity of the system to environmental perturbations. This result provides quantitatively grounded guidance for UAV-borne GPR deployment in cryosphere environments, emphasizing the importance of precise altitude control, trajectory stabilization, and dielectric-property estimation for maintaining coherent imaging performance under harsh environmental conditions.

**Challenges and Future Trends in Power Management, System Miniaturization, and Real-Time 3D Imaging**

Operational constraints also remain a key limiting factor. The endurance of multirotor UAVs is restricted by battery capacity, which becomes particularly critical in low-temperature environments that accelerate power depletion. Maintaining low flight altitudes to ensure data quality further increases power consumption, shortening the effective mission time. Consequently, continuous line coverage and high-density spatial sampling are often difficult to achieve within a single sortie. While autonomous flight planning and waypoint navigation improve safety and consistency—especially in avalanche-prone or crevassed regions—mission efficiency is still largely governed by power management [104]. Field campaigns have shown that careful optimization of flight paths, efficient battery replacement strategies, and on-site recharging infrastructure can partially offset these limitations [104]. However, large-area mapping continues to require multiple power modules, auxiliary energy sources such as portable generators or solar-assisted charging systems, and logistics support for cold-weather maintenance.

Technological progress is expected to gradually alleviate these constraints. Radar subsystem miniaturization, enabled by compact digital transceivers, integrated timing electronics, and efficient antenna matching networks, reduces system mass and improves payload balance [26]. Innovative antenna designs, including conformal or reconfigurable arrays, can enhance polarization control and bandwidth while minimizing aerodynamic drag. In addition, the use of multi-antenna or frequency-stepped arrays offers the potential to increase lateral coverage per sweep and facilitate wide-area 3D reconstruction [91]. From a signal processing perspective, UAV-borne GPR imposes unique challenges due to its contactless geometry and limited angular diversity. The oblique incidence of transmitted waves reduces backscatter strength and alters the effective footprint of the antenna, resulting in incomplete spatial sampling and degraded focusing quality in standard migration algorithms. Therefore, advanced imaging methods that explicitly model platform motion, antenna radiation patterns, and variable propagation velocity are necessary. Emerging approaches have shown promise in reconstructing high-resolution subsurface profiles from sparse or noisy datasets [104], [115], [163]. Extending these frameworks toward real-time, fully three-dimensional reconstruction—potentially executed through onboard processing hardware—will be critical for enabling adaptive survey strategies and efficient data acquisition in dynamic field conditions [24].

Real-time onboard processing remains a fundamental challenge for resource-constrained platforms such as UAVs. Several technological bottlenecks currently limit the realization of fully real-time data processing, imaging and interpretation, including latency constraints in onboard SAR focusing, memory bandwidth limitations during high-rate raw-data streaming, and stringent power-efficiency requirements in embedded computing platforms. Conventional SAR focusing algorithms, such as time-domain back-projection (BP) and frequency-domain methods, require intensive computation for high-resolution and wide-swath imaging, resulting in substantial processing latency that restricts real-time image generation. To address this issue, hardware acceleration techniques—particularly FPGA-based implementations—have been widely investigated to improve computational efficiency and reduce processing latency [191], [192]. These approaches employ reconfigurable processing elements to optimize core SAR computations and enable resource reuse across multiple tasks. In addition, optimized instruction scheduling strategies, including matrix-transpose preprocessing and instruction-caching mechanisms, are introduced to mitigate latency caused by frequent kernel switching [191]. In parallel, algorithm-level optimizations, such as multi-level dataflow parallelism, enable task-level, instruction-level, and node-level concurrency, enable overlapping execution within the processing pipeline, thereby significantly enhancing processing throughput [192].

Meanwhile, the massive data throughput during raw-data streaming—often ranging from hundreds of MB/s to several GB/s—places significant pressure on memory bandwidth and data-transfer efficiency. To mitigate this bottleneck, several strategies have been investigated [193]. From the hardware perspective, techniques such as on-chip cache utilization and unified memory architectures for heterogeneous CPU–GPU systems can reduce reliance on off-chip memory and improve data-transfer efficiency [193]. From the software perspective, optimized memory-access strategies, including data-read optimization and shared-memory access optimization, help reduce global memory accesses while improving data reuse and cache hit rates. In addition, optimized FFT libraries (e.g., cuFFT) can further streamline data movement between on-chip cache and shared memory, thereby reducing memory-access latency and improving overall processing throughput [193].

In addition, power consumption remains a critical limitation for onboard processing systems. Energy-efficient hardware design strategies, including fixed-point implementations, DSP-based parallel acceleration, and multi-stage pipelined architectures, can substantially reduce computational energy consumption while maintaining high throughput [194]. From the perspective of software-level power optimization, lightweight AI inference techniques can be adopted to enable efficient edge processing. Model compression methods, such as network pruning, can reduce model complexity by removing redundant convolutional and fully connected parameters. In addition, quantized neural networks use low-bit representations (e.g., 4-bit or 8-bit) for weights and activations, thereby decreasing computational and memory requirements while maintaining inference performance on embedded platforms. Recent experimental results on the Xilinx ZCU102 FPGA demonstrated a 2.9× speed improvement and a reduction in energy consumption to 68.7% of the original level [195].

Building upon the aforementioned power consumption optimization strategies, further improvements in UAV-borne GPR endurance require advancing the energy supply system beyond generic high energy-density battery enhancements toward application-driven, system-level optimization [196]. Future research should focus on advanced chemistries (e.g., high-nickel Li-ion, Li–S, and solid-state batteries) targeting >300–500 Wh/kg at the pack level, while maintaining sufficient power density for transient flight demands [197]. In parallel, hybrid energy architectures, particularly battery–supercapacitor and fuel cell–battery/SC systems [198], offer a practical pathway to decouple energy and power requirements, with supercapacitors handling peak loads and mitigating rapid battery degradation. Beyond onboard energy improvements, additional approaches to extend operational time include wireless and *in-situ* energy replenishment techniques [199] (e.g., gust soaring, photovoltaic arrays, laser power transfer, and battery swapping/dumping) as well as EM field-based power transfer methods [200], which provide alternative pathways for sustained operation. Overall, these directions highlight a shift from purely chemistry-driven improvements to integrated energy system design, enabling meaningful gains in system endurance and operational reliability for cryosperic scenarios.

Future developments should also emphasize the enhancement of positioning precision and georeferencing accuracy, which are fundamental for coherent multi-pass processing and temporal change detection. The integration of multi-frequency, multi-constellation GNSS-RTK modules enables centimeter-level accuracy at decreasing cost, while sensor fusion with IMU, LiDAR altimeter, or stereo visual odometry can further refine altitude estimation under GNSS-denied conditions. High-precision synchronization between positioning data and radar sampling clocks is equally essential for accurate image reconstruction and repeatability in time-lapse surveys [26], [105].

## CONCLUSIONS

This review synthesized the evolution and applicability of UAV-borne GPR systems in cryosphere environments, highlighting their growing potential for high-resolution, non-invasive shallow subsurface imaging.

In recent years, with continued miniaturization of radar subsystems and the typically low-interference conditions characteristic of cryosphere regions, GPR payloads have been increasingly deployed on small- to medium-scale UAVs for detailed shallow subsurface investigations. UAV-borne GPR has thus emerged as a promising alternative to conventional ground-coupled and other non-contact radar platforms, offering flexible, high-resolution, and cost-effective survey capabilities.

Benefiting from advanced data processing workflows—including preprocessing, reflection and diffraction separation, subsurface velocity analysis, and 2D/3D imaging—UAV-borne GPR has demonstrated effectiveness across diverse cryosphere scenarios, such as snowpack profiling, avalanche rescue operations, ice thickness estimation, and bedrock mapping. By employing CRS-based coherent stacking and subtraction schemes combined with WNE-based velocity analysis, this study further presents a detailed 2D analysis and 3D visualization of UAV-borne GPR data acquired over frozen ice at Mochou Lake in East Antarctica. Comparison with *in-situ* drilling data confirmed the reliability of the approach, achieving an ice thickness estimation error of only 0.03 m.

Although UAV-borne GPR is transitioning from an experimental mapping concept to a relatively mature, high-resolution platform for cryosphere and geotechnical applications, several technical and operational challenges remain. Key issues include signal attenuation, platform stability, georeferencing precision, limited payload capacity and endurance, as well as the complexity of real-time data processing. Addressing these challenges will require coordinated advancements in lightweight antenna and system design, power-efficient flight control, adaptive trajectory planning, and intelligent data processing frameworks. These developments call for synergistic innovation across GPR hardware, algorithmic methodologies, and UAV platform technologies.

Overall, UAV-borne GPR represents a transformative advancement in cryosphere remote sensing study, bridging the gap between localized ground surveys and large-scale airborne or satellite observations. Continued integration of compact radar technologies with autonomous aerial systems and artificial intelligence-driven imaging approaches will further enhance 3D shallow subsurface reconstruction, enabling comprehensive monitoring of cryosphere dynamics in polar and alpine environments.

## Appendix

To ensure the resolution of the final radar image and to avoid aliasing while maintaining measurement accuracy, we need to consider the Doppler bandwidth $f_{dop}$ [201], [202]. Specifically, the image resolution is dictated by the Doppler support range, a direct consequence of the Doppler shifts caused by the relative sensor-target motions.

The Doppler frequency is written as

$$f_{dop} = -\frac{2}{\lambda}\cdot\frac{\partial Z(\eta)}{\partial \eta}, \qquad (A-1)$$

where $\eta$ is time, $Z$ is the distance between the UAV and the subsurface target.

The approximation for $Z(\eta)$ is

$$Z(\eta) = \sqrt{{Z_0}^2 + {v_r}^2\eta^2}, \qquad (A-2)$$

Then, the derivative of $Z(\eta)$ with respect to $\eta$ is:

$$\frac{\partial Z(\eta)}{\partial \eta} = \frac{2{v_r}^2\eta}{2\sqrt{{Z_0}^2 + {v_r}^2\eta^2}} = \frac{{v_r}^2\eta}{Z(\eta)}, \qquad (A-3)$$

Thus, the Doppler frequency $f_{dop}$ becomes

$$f_{dop}(\eta) = -\frac{2v_r}{\lambda}\cdot\frac{v_r\eta}{Z(\eta)}. \qquad (A-4)$$

In terms of the down-looking UAV-borne GPR system, the Doppler bandwidth can be expressed as

$$\Delta f_{dop} = max\left(f_{dop}(\eta)\right) - min\left(f_{dop}(\eta)\right). \qquad (A-5)$$

When $\eta = \pm(L_s/2v_r)$ , we obtain the maximum and minimum values of $f_{dop}$. The Doppler bandwidth can then be rewritten as

$$\Delta f_{dop} = \frac{2v_r}{\lambda}\frac{L_s}{Z_0}, \qquad (A-6)$$

where

$$Z_0 = \frac{(L_s/2)}{sin(\theta_{3dB}/2)}. \qquad (A-7)$$

Finally, the Doppler bandwidth can be written as

$$\Delta f_{dop} = \frac{4v_r}{\lambda} sin(\theta_{3dB}/2). \qquad (A-8)$$

To ensure the resolution of the migrated image, avoiding aliasing and maintaining measurement accuracy, the PRF, which is defined as $PRF = v_r num/L_s$, must be greater than $\Delta f_{dop}$ as

$$PRF = \frac{v_r num}{L_s} \geq \Delta f_{dop} = \frac{4v_r}{\lambda} sin(\theta_{3dB}/2), \qquad (A-9)$$

which leads to (17). Thus,

$$num \geq \frac{4L_s}{\lambda} sin(\theta_{3dB}/2), \qquad (A-10)$$

where $num$ is the number of A-scan points in $L_s$ range:

$$L_s = 2R_0\, tan(\theta_{3dB}/2), \qquad (A-11)$$

where $L_s$ is determined by $R_0$ when the GPR antenna beamwidth ($\theta_{3dB}$) is fixed.

Finally, we get (18) as

$$n \geq \frac{8Z_0\, tan\left(\frac{\theta_{3dB}}{2}\right)}{\lambda} sin\left(\frac{\theta_{3dB}}{2}\right). \qquad (A-12)$$

## Acknowledgements

Constructive comments from the editors and reviewers are greatly appreciated. We thank the 36th CHINARE team and the Polar Research Institute of China for logistic and technical support. The support by Dalian Zoroy Technology Development Co., Ltd. is appreciated. This study was supported by the National Key Research and Development Program of China (grant 2021YFB3900105), the National Natural Science Foundation of China (grants 42504145, 42574201 and 42074179), and Fundamental Research Funds for the Central Universities. Tong Hao is the corresponding author.

## Author Information

**Wenhao Luo** (wenhao_luo@tongji.edu.cn) received the B.Eng. degree in automation from Hohai University, Nanjing, China, in 2016, the M.Sc. and Ph.D. degrees in electrical and electronic engineering from the School of Electrical and Electronics Engineering, Nanyang Technological University, Singapore, in 2019 and 2023, respectively. From 2023 to 2026, he was a Post-Doctoral Research Fellow with Tongji University, Shanghai 200092, China, where he is currently an Assistant Professor with the College of Surveying and Geo-Informatics. His research focuses on ground-penetrating radar applications for civil infrastructure and the cryosphere, as well as remote sensing, signal processing, and machine learning.

**Tong Hao** (tonghao@tongji.edu.cn) received the B.Sc. degree from Nanjing University, Nanjing, China, in 2003, the M.Phil. degree in electronic and electrical engineering from the University of Bath, Bath, U.K., in 2004, and the D.Phil. degree in engineering science from the University of Oxford, Oxford, U.K., in 2009. From 2010 to 2011, he was a Postdoctoral Research Fellow with the University of Birmingham, Birmingham, U.K. From 2012 to 2014, he was a Specialist with the Oil and Gas Sector, General Electric Company, Farnborough, U.K. He is currently a Full Professor with the College of Surveying and Geo-Informatics, Tongji University, Shanghai 200092, China. His research interests include enhanced electromagnetic detection and sensing of subsurface targets, and wireless sensor networks in complex media.

**Qian Ma** (qianma@tongji.edu.cn) is currently pursuing the Ph.D. degree at Tongji University, Shanghai 200092, China, with the Shanghai Research Institute for Intelligent Autonomous Systems. Her research interests include polar radar remote sensing, deep-learning-based radar image analysis, and subglacial topographic modeling.

**Chen Lv** (carolina@tongji.edu.cn) received the B.Eng. degree from the College of Surveying and Geo-Informatics at Tongji University, Shanghai 200092, China, in 2021, where she is currently pursuing the Ph.D. degree. Her research interests include subsurface sounding radar, remote sensing techniques, and signal processing and interpretation in radioglaciology.

**Zhiyi Cao** (caozhiyi@tongji.edu.cn) received the B.Eng. degree from the College of Surveying and Geo-Informatics at Tongji University, Shanghai 200092, China, in 2024, where he is currently pursuing the Ph.D. degree. His research interests include ground penetrating radar and subsurface targets reconstruction.